\documentclass{elsart}
\usepackage{ragged2e}
\usepackage{graphics}
\usepackage{graphicx}
\usepackage{epsfig}
\usepackage{amssymb}
\usepackage{lineno}
\usepackage{blindtext}
\usepackage{mathtools}
\begin{document}
\begin{frontmatter}
\title{Critical corrections to formulations of nonlinear energy dissipation of ultrasonically excited bubbles and a unifying parameter to asses and enhance bubble activity in applications}
\author{A.J. Sojahrood \thanksref{a},\thanksref{b}} \footnote{Email: amin.jafarisojahrood@ryerson.ca},
\author{H. Haghi \thanksref{a},\thanksref{b}}
\author{R. Karshfian \thanksref{a},\thanksref{b}}
\author{and  M. C. Kolios \thanksref{a},\thanksref{b}}
\address[a]{Department of Physics, Ryerson University, Toronto, Canada}
\address[b]{Institute for Biomedical Engineering, Science and Technology (IBEST) a partnership between Ryerson University and St. Mike's Hospital, Toronto, Ontario, Canada}
\begin{abstract}
	Current models for calculating nonlinear energy dissipation in the oscillation of acoustically excited bubbles generate non-physical values for the radiation damping term for some frequency and pressure regions including near resonance oscillations. We provide critical corrections to the present formulations. We highlighted the importance of the  corrections by calculating the scattering to damping ratio (STDR) for nonlinear regime of oscillations. We then introduce two new parameters to assess the efficacy of applications. They are defined as the multiplication of maximum re-radiated pressure and radiation dissipation by STDR.
\end{abstract}
\end{frontmatter}

\section{Introduction}
Bubbles attenuate ultrasound through viscous damping due to liquid friction (Ld), radiation damping (Rd) and thermal damping (Td) \cite{1,2,3,4,5,6}. Numerous studies have investigated the mechanisms of damping in bubbly media; using linear  approximations (limited to very small bubble oscillation amplitudes) and neglecting the dependence of the dissipated energy on the local pressure \cite{4,5,6,7,8}. Semi-linear approaches have also been developed that only considered the pressure dependance of the radiation damping while using linear terms for the other damping factors (thermal and viscous damping) \cite{9}. Additionally, the role of Td is simplified using models that are derived based on linear approximations \cite{4,5,6,7,8,9}. A more complete estimation of the wave attenuation in bubbly media requires a realistic estimation of the power dissipated by the nonlinear oscillations of the bubbles that includes all forms of damping \cite{1,2,3,10,11}.\\
Louisnard \cite{1}, starting with mass and momentum conservation equations for a bubbly liquid and using Rayleigh-Plesset equation \cite{12} have derived the nonlinear energy terms for Td and Ld. He showed that damping from nonlinear oscillations of the bubbles can be several orders of magnitude higher than the damping estimated by linear models. Jamshidi and Brenner \cite{2} used the approach introduced by Louisnard \cite{1} in conjunction  with the Keller-Miksis equation \cite{13} accounting for the compressibility of the medium to the first order of acoustical Mach number. Incorporation of the changes in the compressibility of the medium allowed for derivation of Rd and small modifications in Td and dL. It was shown \cite{2,3} that radiation damping increases significantly above Blake threshold and becomes one of the major contributors to total damping and cannot be neglected. However, as it will be shown here, the terms that are derived in \cite{2}, have errors and need to be properly derived. Rd at its current form \cite{2} leads to negative values near resonance and in some frequency and pressure ranges. 
A damping factor should always have positive values; it can not be negative. In this work, nonlinear terms for Td, Rd and Ld were re-arranged. Using the pressure radiated by pulsating bubbles \cite{4,14,15,16} the accuracy of the new
formulation is verified.\\
Using the newly derived terms, nonlinear scattering to damping ratio (STDR) is calculated for nonlinear oscillation regimes of interest. Two new nonlinear parameters are  proposed as a unifying factor for assessing the applications efficacy. The first parameter (PmSTDR) is defined as the multiplication of maximum scattered pressure ($m_P$) by STDR. The second parameter is defined as the multiplication of Rd with STDR.  Using the formulation presented here, PmSTDR can be calculated accurately and used for optimization of the parameters of  applications for enhanced outcome.

\section{Methods}
\subsection{Mass and momentum equations for bubbly media}
van Wijngaardan \cite{17} and Caflish et al. \cite{18}
presented the mass and momentum conservation equations for a
bubbly liquid as:
\begin{equation}
\frac{1}{\rho c^2}\frac{\partial P}{\partial t}+\nabla.v=\frac{\partial \beta}{\partial t}
\end{equation}
and
\begin{equation}
\rho\frac{\partial v}{\partial t}=-\nabla P
\end{equation}
where $c$ is the sound speed, $\rho$ is the density of the medium, $v(r,t)$ is the velocity field, $P(r,t)$ is acoustic pressure, $\beta=\frac{4}{3}N\pi R(t)^3$ is the void fraction where N is number of bubbles per unit volume, and $R(t)$ is the radius of the bubble at time $t$. These two equations can be re-written into an equation of energy conservation, by multiplying (1) by $P$ and (2) by $v$:
\begin{equation}
\frac{\partial}{\partial t}\left(\frac{1}{2}\frac{P^2}{\rho c^2}+\frac{1}{2}\rho v^2\right)=NP\frac{\partial V}{\partial t}
\end{equation}

\subsection{The Bubble model}
The dynamics of the bubble model including the compressibility effects to the first order of Mach number can be modelled using Keller-Miksis equation\cite{13}:
\justifying
\begin{equation}
\rho[(1-\frac{\dot{R}}{c})R\ddot{R}+\frac{3}{2}\dot{R}^2(1-\frac{\dot{R}}{3c})]=(1+\frac{\dot{R}}{c})(G)+\frac{R}{c}\frac{d}{dt}(G)
\end{equation}
where $G=P_g-\frac{4\mu_L\dot{R}}{R}-\frac{2\sigma}{R}-P_0-P_A sin(2 \pi f t)$.\\
In this equation, R is radius at time t, $R_0$ is the initial bubble radius, $\dot{R}$ is the wall velocity of the bubble, $\ddot{R}$ is the wall acceleration,	$\rho{}$ is the liquid density (998 $\frac{kg}{m^3}$), c is the sound speed of the medium (1481 m/s), $P_g$ is the gas pressure, $\sigma{}$ is the surface tension (0.0725 $\frac{N}{m}$), $\mu{}$ is the liquid viscosity (0.001 Pa.s), and $P_A$ and \textit{f} are the amplitude and frequency of the applied acoustic pressure. The values in the parentheses are for pure water at 293$^0$K. In this paper the gas inside the bubble is air and water is the host media.\\

$P_g$ is given by Eq. 5 \cite{19}:
\begin{equation}
P_g=\frac{N_gKT}{\frac{4}{3}\pi R(t)^3-NB}
\end{equation}
Where $N_g$ is the total number of the gas molecules, $K$ is the Boltzman constant and B is the molecular co-volume. The average temperature inside the gas can be calculated using Eq. 6:
\begin{equation}
\dot{T}=\frac{4\pi R(t)^2}{C_v} (\frac{L(T_0-T)}{L_{th}}-\dot{R}P_g)
\end{equation}
Where $C_v$ is the specific heat at constant volume, $T_0$=$293^0$K is the initial gas temperature, $L_{th}$ is the thickness of the thermal boundary layer. $L_{th}$ is given by $L_{th}=min(\sqrt{\frac{aR(t)}{|\dot{R(t)}|}},\frac{R(t)}{\pi})$ where $a$ is the thermal diffusivity of the gas. $a$ can be calculated using $a=\frac{L}{C_p \rho_g}$ where L is the gas thermal conductivity and $C_p$ is specific heat at constant pressure and $\rho_g$ is the gas density. 

\begin{table}
	\begin{tabular}{ |p{3.5cm}|p{2cm}|p{2cm}|p{2cm}|  }
		\hline
		\multicolumn{4}{|c|}{Thermal parameters of the Air at 1 atm \cite{20}} \\
		\hline
		L ($\frac{W}{m^0K}$) &$C_p$$\frac{kJ}{kg^0C}$ &$C_v$ $\frac{kJ}{kg^0C}$&$\rho_g$ $\frac{kg}{m^3}$\\
		\hline
		0.01165+C*T  &1.0049&   0.7187&1.025\\
		\hline
	\end{tabular}
	\caption{Thermal properties used in simulations. (C=$5.528*10^{25}$ $\frac{W}{m^0K^2}$)} 
	\label{table:1}
\end{table}

To calculate the radial oscillations of the bubble Eqs. 4, 5 and 6 are coupled and solved using the ode45 solver of Matlab.
\subsection{Derivation of the damping terms}
Multiplying both sides of Eq.4 by $\frac{N\partial V}{\partial t}$ and summation with equation 3 yields:

$$\rho N\left(R\ddot{R}+\frac{3}{2}\dot{R}^2\right)\frac{\partial V}{\partial t}-\rho N\frac{\dot{R}}{c}\left(R\ddot{R}+\frac{1}{2}\dot{R}^2\right)\frac{\partial V}{\partial t}+\frac{\partial}{\partial t}\left(\frac{1}{2}\frac{P^2}{\rho c^2}+\frac{1}{2}\rho v^2\right)+\nabla.(Pv)$$
\\
$$=N\left(P_g+\frac{\dot{R}}{c}P_g+\frac{R}{c}\frac{dP_g}{dt}\right)\frac{\partial V}{\partial t}-N\left(\frac{4\mu_L \dot{R}}{R}+\frac{\dot{R}}{c}\frac{4\mu_L \dot{R}}{R}+\frac{R}{c}\frac{d(\frac{4\mu_L \dot{R}}{R})}{dt}\right)\frac{\partial V}{\partial t} $$
\begin{equation}
-N\left(\frac{2\sigma}{R}+\frac{\dot{R}}{c}\frac{2\sigma}{R}+\frac{R}{c}\frac{d(\frac{2\sigma}{R}}{dt}\right)\frac{\partial V}{\partial t}
-N\left(\frac{\dot{R}}{c}P+\frac{R}{c}\frac{dP}{dt}\right)\frac{\partial V}{\partial t}
\end{equation}
The kinetic energy of the liquid around bubble can be written as \cite{2}:
\begin{equation}
K_l=2\pi \rho R^3 \dot{R}^2
\end{equation}
using equation 7 and 8 and re-arranging terms we will have:

$$
\frac{\partial}{\partial t}\left(\frac{1}{2}\frac{P^2}{\rho c^2}+\frac{1}{2}\rho v^2+NK_l+4N\pi R^2 \sigma\right)+\nabla.(Pv)=N\left(P_g\right)\frac{\partial V}{\partial t}-N\left(\frac{4\mu_L\dot{R}}{R}\right)-$$
\\
\begin{equation}
N\left(\left(\frac{\dot{R}}{c}P+\frac{R}{c}\frac{dP}{dt}-\frac{\dot{R}}{c}P_g-\frac{R}{c}\frac{dP_g}{dt}+\frac{\dot{R}}{c}\frac{4\mu_L\dot{R}}{c}+\frac{R}{c}\frac{d(\frac{4\mu_L\dot{R}}{R})}{dt}\right)\frac{\partial V}{\partial t}-\frac{\dot{R}}{c}\frac{\partial K_l}{\partial t}\right)
\end{equation}

Eq.9 can be written as:

\begin{equation}
\begin{gathered}
\frac{\partial}{\partial t}\left(\frac{1}{2}\frac{P^2}{\rho c^2}+\frac{1}{2}\rho v^2+NK_l+4N\pi R^2 \sigma\right)+\nabla.(Pv)=-N\left(td+ld+rd\right)
\end{gathered}
\end{equation}	
where td, ld and rd are time dependent thermal, liquid and radiation damping respectively. For N=1, Jamshidi and Brenner presented the damping terms in Eq. 9 as:
\begin{equation}
\begin{gathered}
\begin{dcases}
td=(-P_g -\frac{\dot{R}}{c}P_g-\frac{R}{c}\dot{P_g})\frac{\partial V}{\partial t}\\
ld=\left(\frac{4\mu_L\dot{R}}{R}+\frac{\dot{R}}{c}\frac{4\mu_L\dot{R}}{c}+\frac{R}{c}\frac{d(\frac{4\mu_L\dot{R}}{R})}{dt}\right)\frac{\partial V}{\partial t}\\
rd=\left(\frac{\dot{R}}{c}P+\frac{R}{c}\frac{dP}{dt}+\right)\frac{\partial V}{\partial t}-\frac{\dot{R}}{c}\frac{\partial K_l}{\partial t}
\end{dcases}
\end{gathered}
\end{equation}
Integrating Eq.11 over one acoustic period $T$ results in:
\begin{equation}
\begin{gathered}
\frac{1}{T}\int_{0}^{T}\frac{\partial}{\partial t}\left(\frac{1}{2}\frac{P^2}{\rho c^2}+\frac{1}{2}\rho v^2+K_l+4\pi R^2\sigma\right)+\nabla.<Pv>=-\left(Td+Ld+Rd\right)
\end{gathered}
\end{equation}
The first term on the left hand side of Eq. 12 cancels over one acoustic period \cite{1,2}. and the total energy loss becomes:
\begin{equation}
\begin{dcases}
Td=\frac{-1}{T}\int_{0}^{T}(P_g +\frac{\dot{R}}{c}P_g+\frac{R}{c}\dot{P}_g)\frac{\partial V}{\partial t}dt\\ \\
Ld=\frac{16\pi\mu_L}{T}\int_{0}^{T}(R\dot{R}^2+\frac{R^2\dot{R}\ddot{R}}{c})dt\\ \\
\begin{gathered}
Rd=\frac{1}{T}\int_{0}^{T} \left[\frac{4\pi}{c}\left(R^2\dot{R}\left(\dot{R}P+R\dot{P}-\frac{1}{2}\rho \dot{R}^3-\rho R\dot{R}\ddot{R}\right)\right)\right]dt
\end{gathered}
\end{dcases}
\end{equation}
where Td, Ld and Rd are total energy loss over time T. However, Eq. 13 in its current format is not correct and several terms need to be re-arranged. Every term that contains the sound speed represents the compressibility effects to the first order of Mach number and should be added to the radiation damping term. Thus the corrected damping terms will be in the form of Eq. 14:
\begin{equation}
\begin{dcases}
Td=\frac{-1}{T}\int_{0}^{T}\left(P_g\right)\frac{\partial V}{\partial t}dt\\ \\
Ld=\frac{16\pi\mu_L}{T}\int_{0}^{T}\left(R\dot{R}^2\right)dt\\ \\
\begin{gathered}
Rd=\frac{1}{T}\int_{0}^{T} \left[\frac{4\pi}{c}\left(R^2\dot{R}\left(\dot{R}P+R\dot{P}-\frac{1}{2}\rho \dot{R}^3-\rho R\dot{R}\ddot{R}\right)\right)\right.\\
\left.-\left(\frac{\dot{R}}{c}P_g+\frac{R}{c}\dot{P}_g\right)\frac{\partial V}{\partial t}+\frac{16\pi\mu_LR^2\dot{R}\ddot{R}}{c}\right]dt
\end{gathered}
\end{dcases}
\end{equation}
\subsection{Acoustic power due to scattered pressure by bubbles}
Radiation damping is due to the re-radiated (scattered) pressure by the bubble.
The acoustic energy scattered by an oscillating bubble can be calculated using \cite{4}:
\begin{equation}
W_{sc}=\frac{4\pi r^2}{\rho c}P_{sc}^2
\end{equation}
where $P_{sc}$ is the pressure scattered (re-radiated) by the oscillating bubble \cite{14,15,16}:
\begin{equation}
P_{sc}=\rho\frac{R}{r}(R\ddot{R}+2\dot{R}^2)
\end{equation}
here $r$ is the distance from the bubble center. Using Eq.15 and Eq.16 we can write:
\begin{equation}
W_{sc}=\frac{4\pi\rho}{c}R^2\left(R\ddot{R}+2\dot{R}^2\right)^2
\end{equation}
The dissipated energy due to radiation should have the same value of the acoustic scattered energy by the bubble. Therefore, one can compare Rd and $W_{sc}$ to validate the predictions of Eq. 14. To demonstrate the regions where Eq. 13 fails to predict correct values, Eq. 4, 5 and 6 were solved over 60 cycles of driving force, then Eq. 13, Eq. 14 and Eq. 17 were used to find the dissipated power for the last 20 cycles of oscillations.
\begin{figure*}
	\begin{center}
		\scalebox{0.33}{\includegraphics{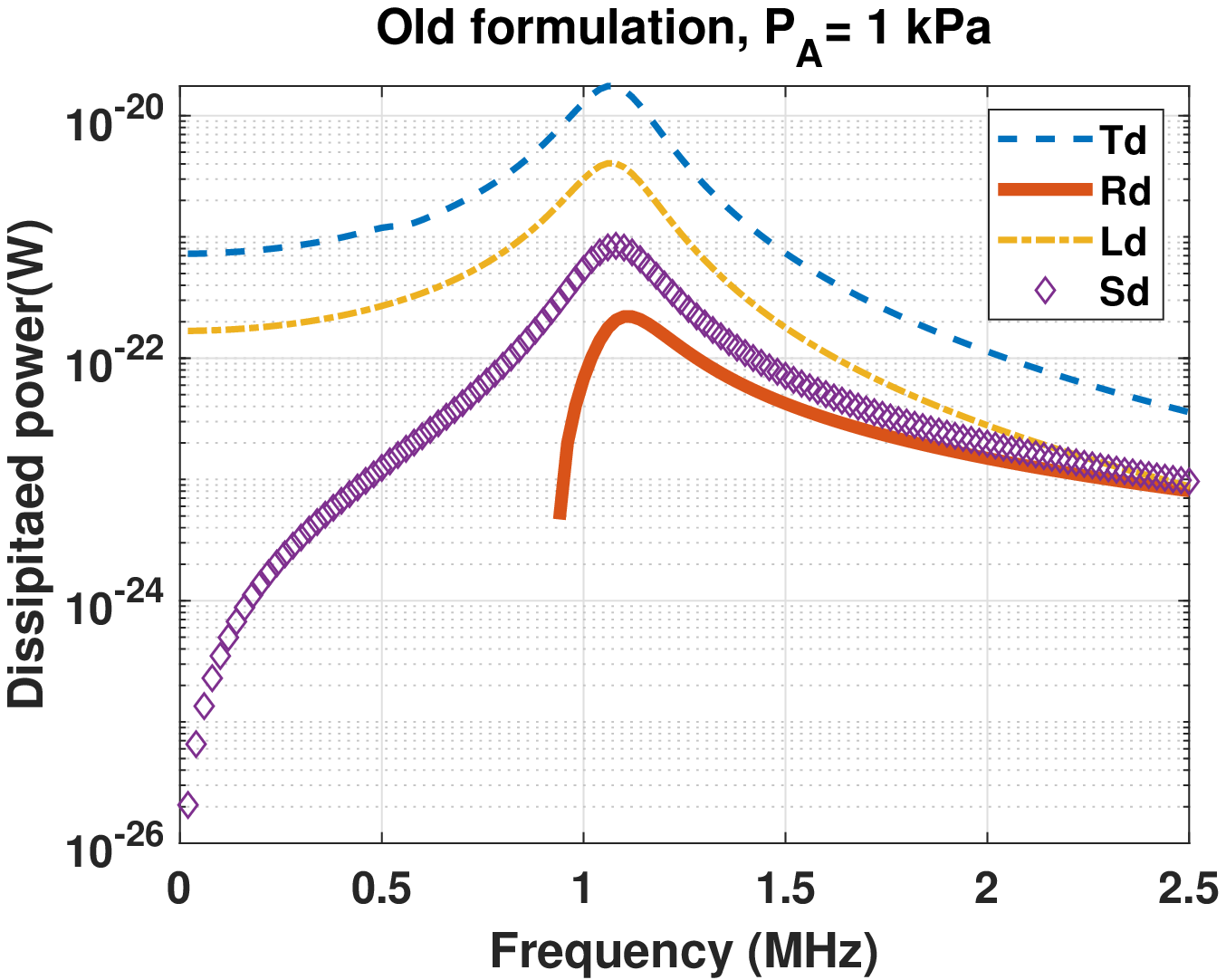}} \scalebox{0.33}{\includegraphics{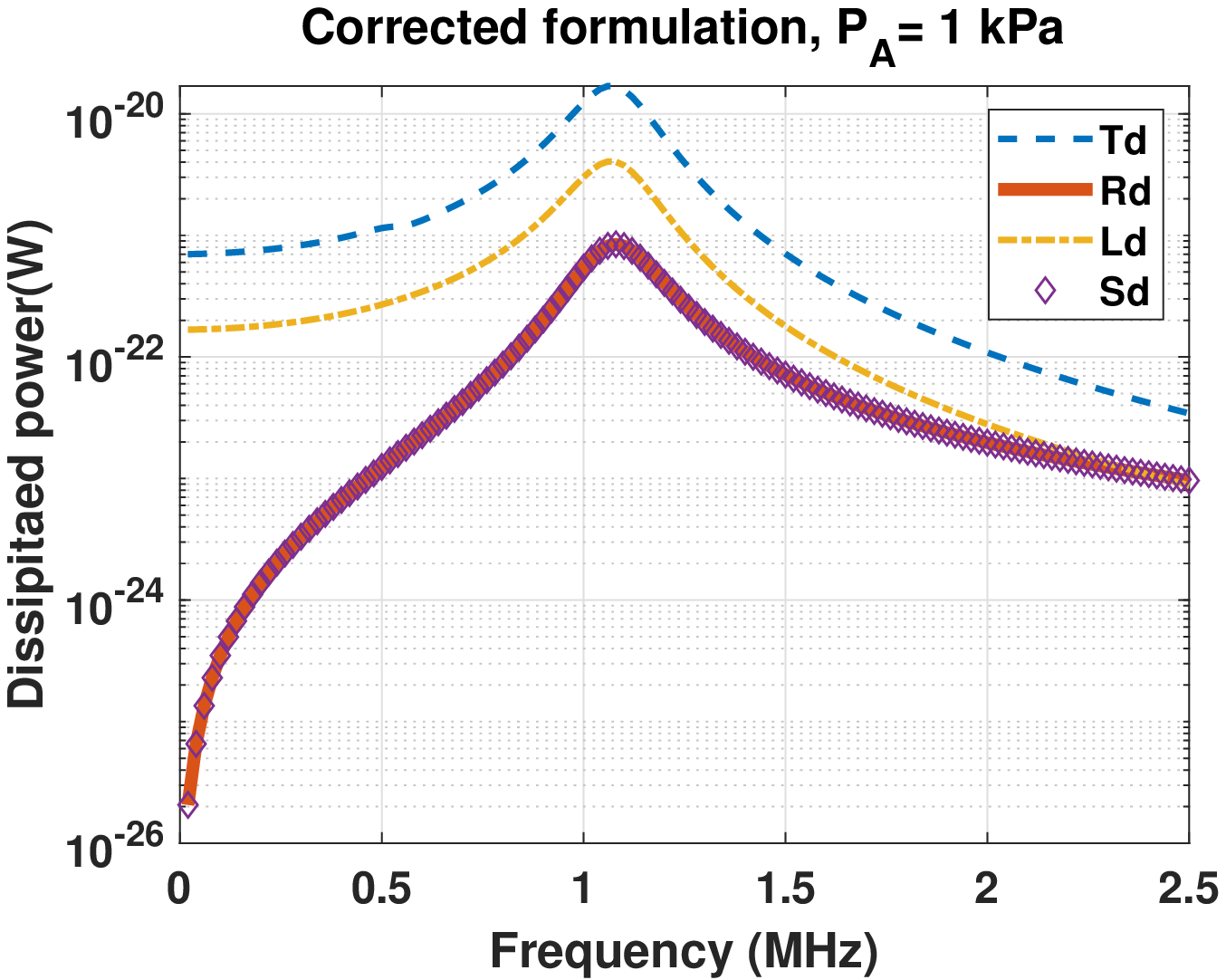}}\\
		\hspace{0.5cm} (a) \hspace{6cm} (b)\\
		\scalebox{0.33}{\includegraphics{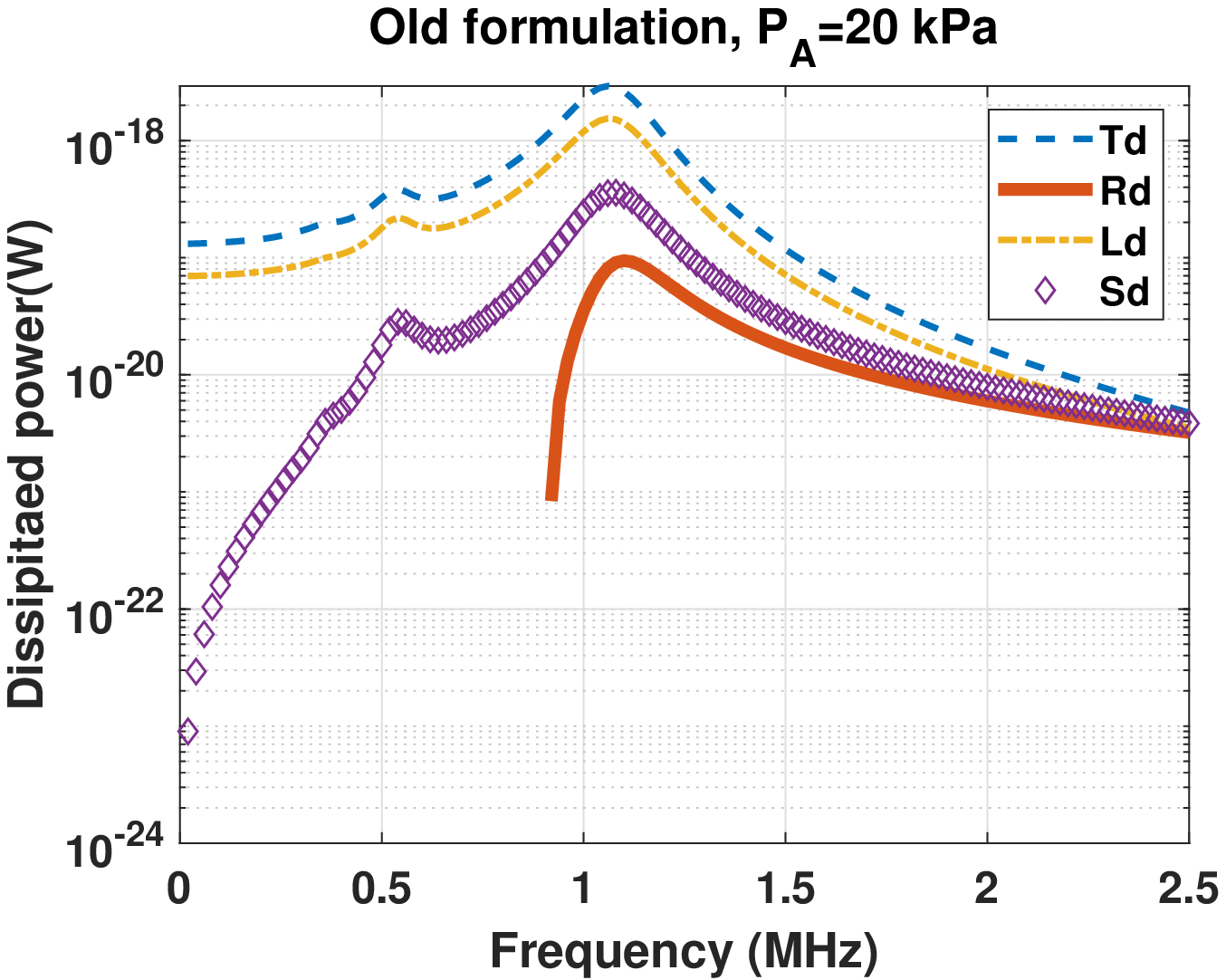}} \scalebox{0.33}{\includegraphics{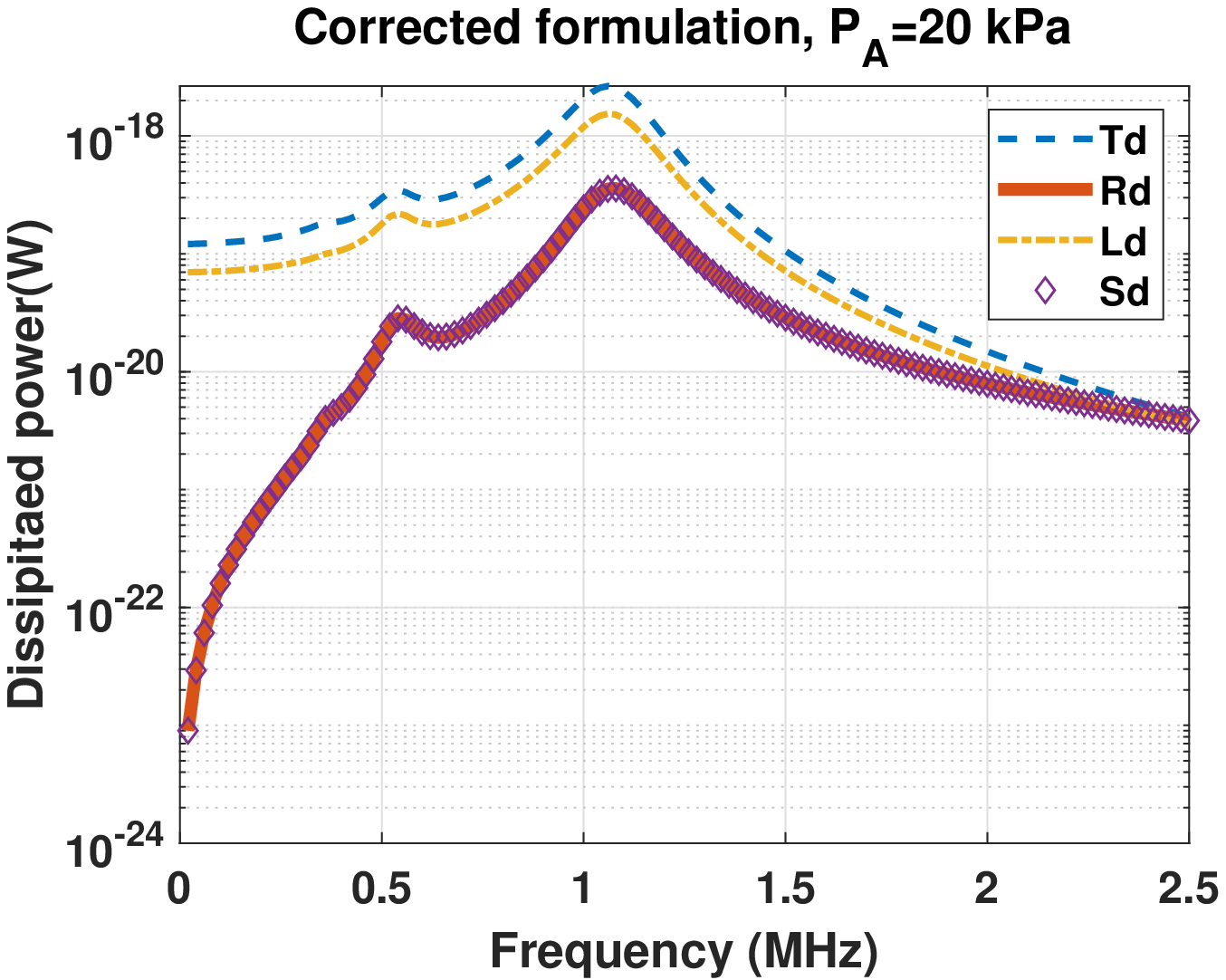}}\\
		\hspace{0.5cm} (c) \hspace{6cm} (d)\\
		\scalebox{0.33}{\includegraphics{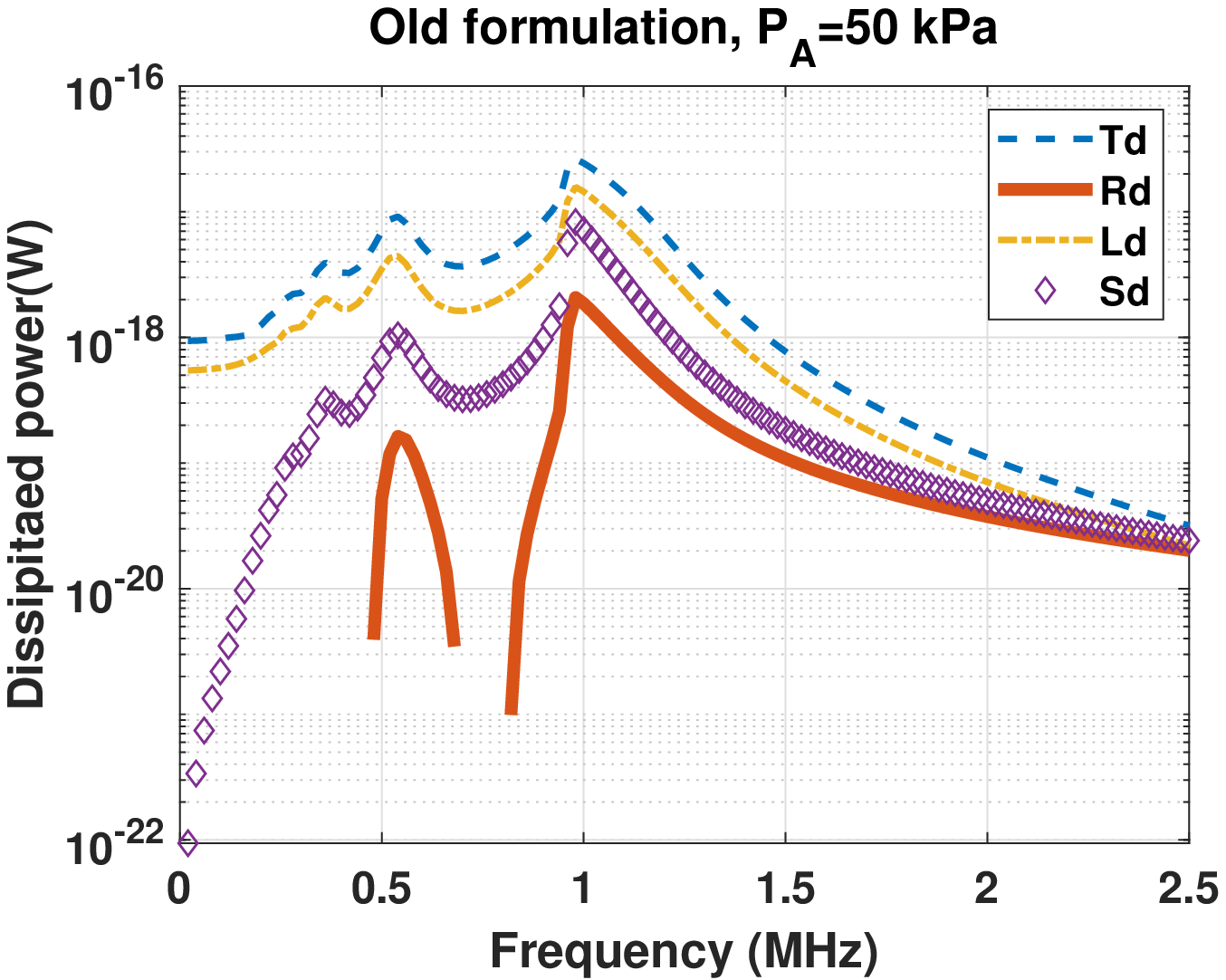}} \scalebox{0.33}{\includegraphics{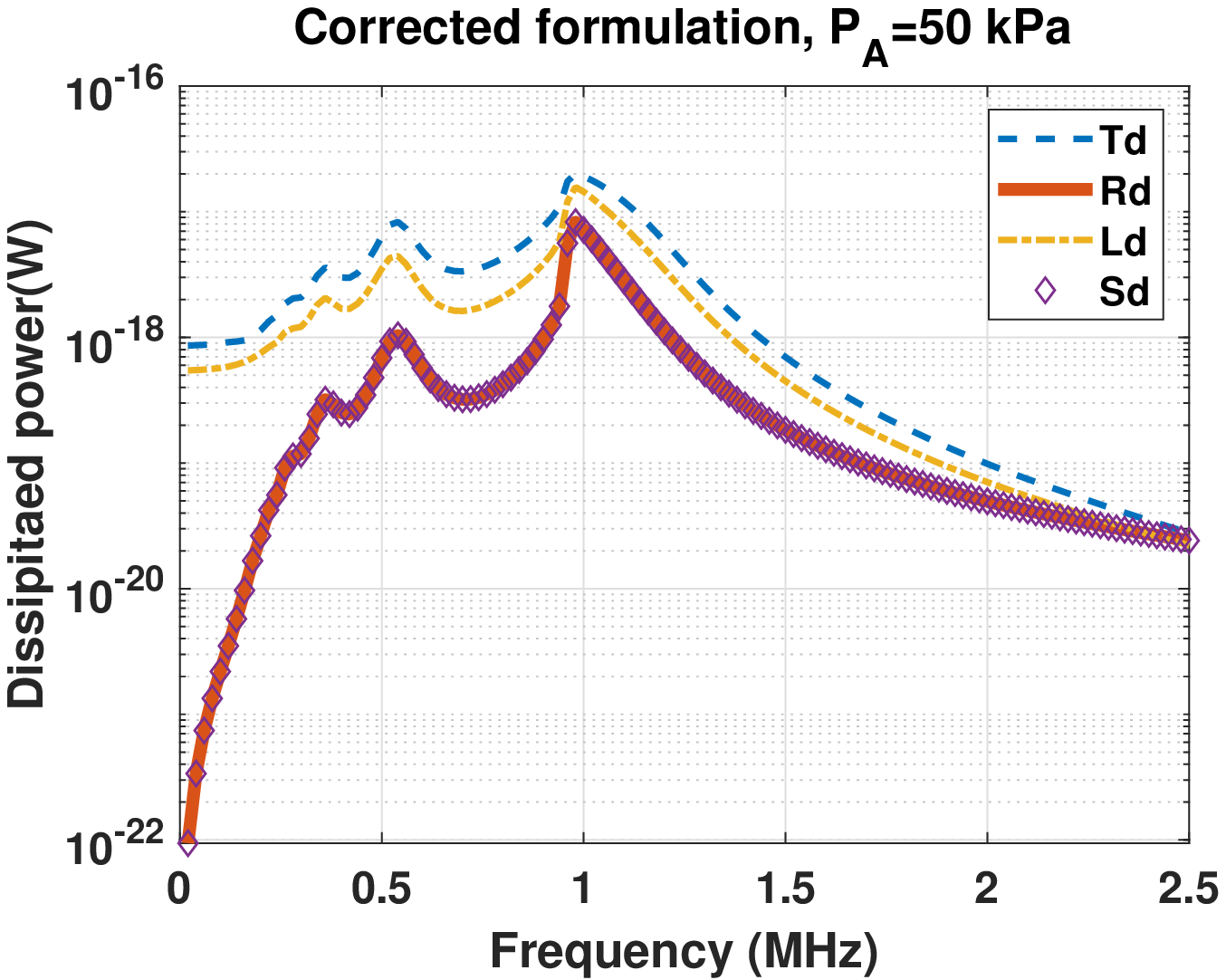}}\\
		\hspace{0.5cm} (e) \hspace{6cm} (f)\\
		\scalebox{0.33}{\includegraphics{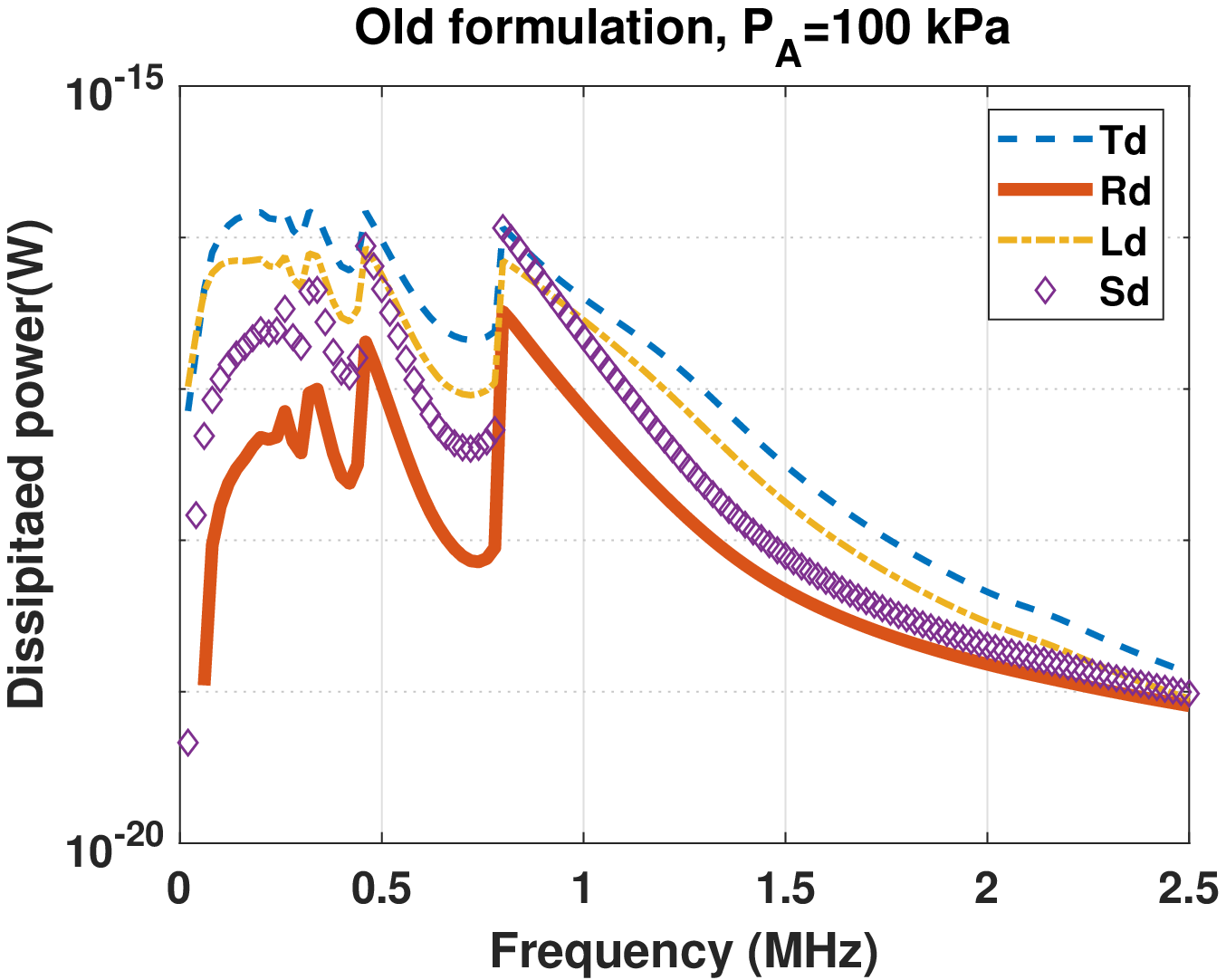}} \scalebox{0.33}{\includegraphics{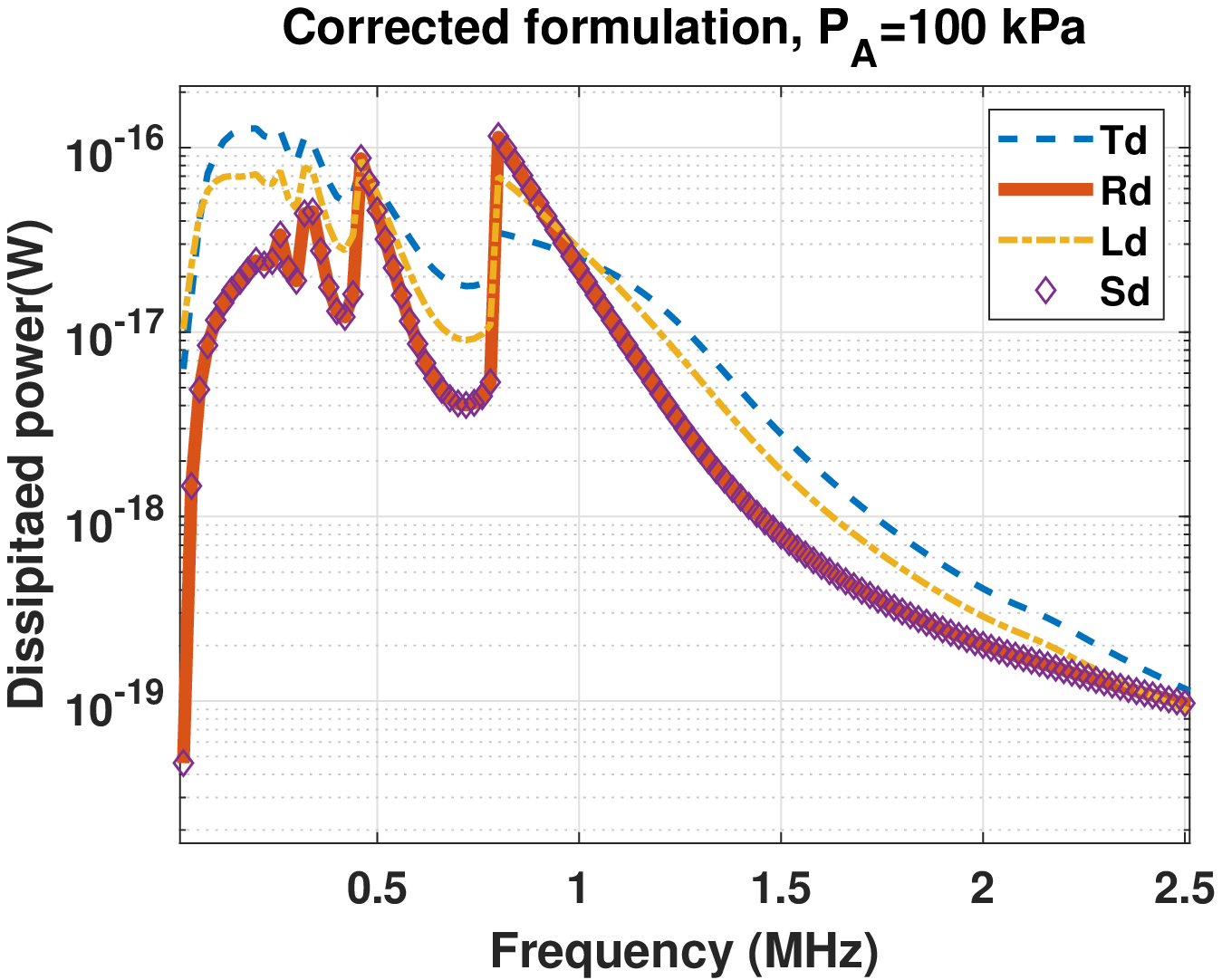}}\\
		\hspace{0.5cm} (g) \hspace{6cm} (h)\\
		\caption{The dissipated power due to Td, Rd, Ld and damping due to scattered pressure by bubble (Sd) when $P_A$=1 kPa as a function of frequency for an air bubble with $R_0$= 3 $\mu$ m estimated by the Jamshidi and Brenner model [2] (left column) and the corrected model (right column).}
	\end{center}
\end{figure*}
\section{Results}
Fig. 1 shows the dissipated power due to thermal damping (Td), liquid viscous damping(Ld), radiation damping (Rd) and damping due to the scattered pressure by the bubble ($W_{sc}$) for acoustic pressures of 1, 20, 50 $\&$ 100 kPa and as a function of frequency for an air bubble with $R_0$=3 $\mu$m. The y-axis is in log format for a better comparison. Figure 1a, represents the predictions made by the old formulations (Eq. 13) when $P_A$=1kPa. Raditaion damping is not equal to $W_{sc}$ and below resonance Rd is negative (negative values are absent in the log graph). Thus, predictions of Eq. 13 are inaccurate. Figure 1b, displays the predictions of the corrected equations (Eq. 14). Rd is exactly equal to $W_{sc}$ and negative values are absent. In addition, at 1 kPa predictions are consistent  with the estimations by the linear model \cite{4,5}.\\
At higher pressures, predictions of the non-linear model significantly deviate from the linear model \cite{1,2,3}. Figures 1c-d shows the dissipated power at $P_A$= 20 kPa due to Td, Ld, Rd and $W_{sc}$ calculated using Eq. 13 and Eq. 14, respectively. For frequencies below resonance, Rd becomes negative and a negative value for a damping mechanism is not possible. A negative value for damping is analogous to a phenomenon that pumps energy to the system. Fig. 1d shows that the corrected model described by Eq. 14 predicts the correct values for Rd which are consistent with $W_{sc}$. We also see the generation of 2nd harmonic resonance frequency due to the pressure increase at $\approx$ 540 kHz.\\ Figure 1e-f, respectively represents the predictions of Eq. 13 and Eq.14 when $P_A$=50 kPa. Increasing the acoustic pressure shifted the fundamental frequency of the bubble to lower values. It increased the damping values by an order of magnitude. Moreover,  a 3rh harmonic resonance peak at ~360 kHz is generated. Once again, Eq. 13 predicts negative values for Rd for some frequencies below the fundamental resonance frequency and there is discrepancy between $W_{sc}$ and the Rd. Eq. 14 accurately captures the value of Rd and there is no negative value present for the damping terms. Figures 1g-h show the predictions of Eq. 13 and 14 when $P_A$= 100 kPa. Fig. 1g shows that solutions of Eq. 13 for Rd are inconsistent with $W_{sc}$ and Rd is negative at 20 kHz; However, the corrected model (Eq. 14) predicts the correct value for Rd in good agreement with $W_{sc}$. Elevating pressure to $P_A$= 100 kPa results in further shifting of the resonance frequencies of the system to lower frequencies, and 4th and 5th harmonic resonance peaks are observable in the graphs. Radiation damping grows faster than the other damping factors as pressure increases. At the pressure dependent resonance (f=800 kHz) and pressure dependent 2nd harmonic resonance (f=460 kHz) radiation damping becomes stronger than the rest of the damping factors. Thermal damping remains the strongest damping term for the rest of frequencies.\\   
\begin{figure*}
	\begin{center}
		\scalebox{0.33}{\includegraphics{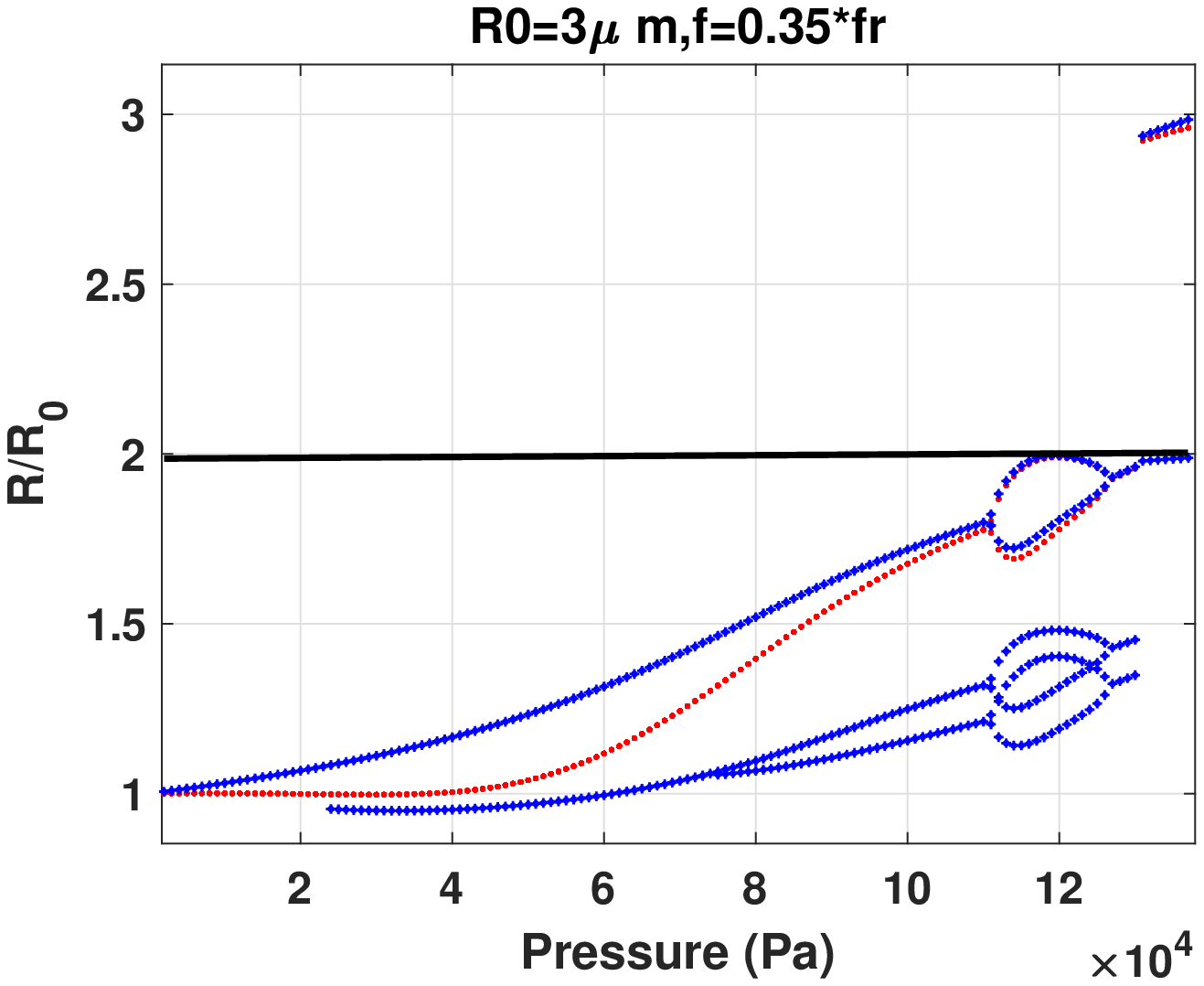}} \scalebox{0.33}{\includegraphics{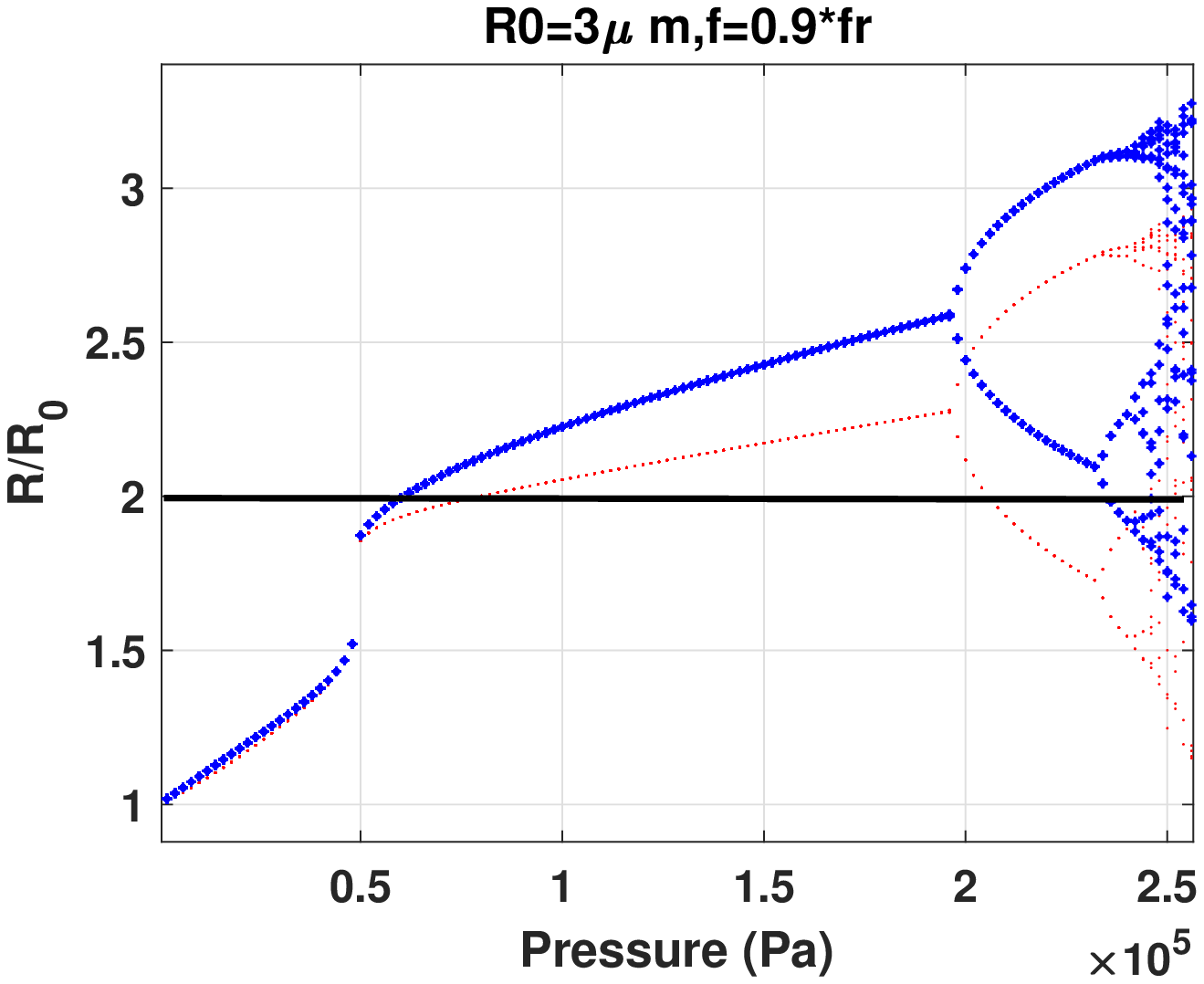}}\\
		\hspace{0.5cm} (a) \hspace{6cm} (b)\\
		\scalebox{0.33}{\includegraphics{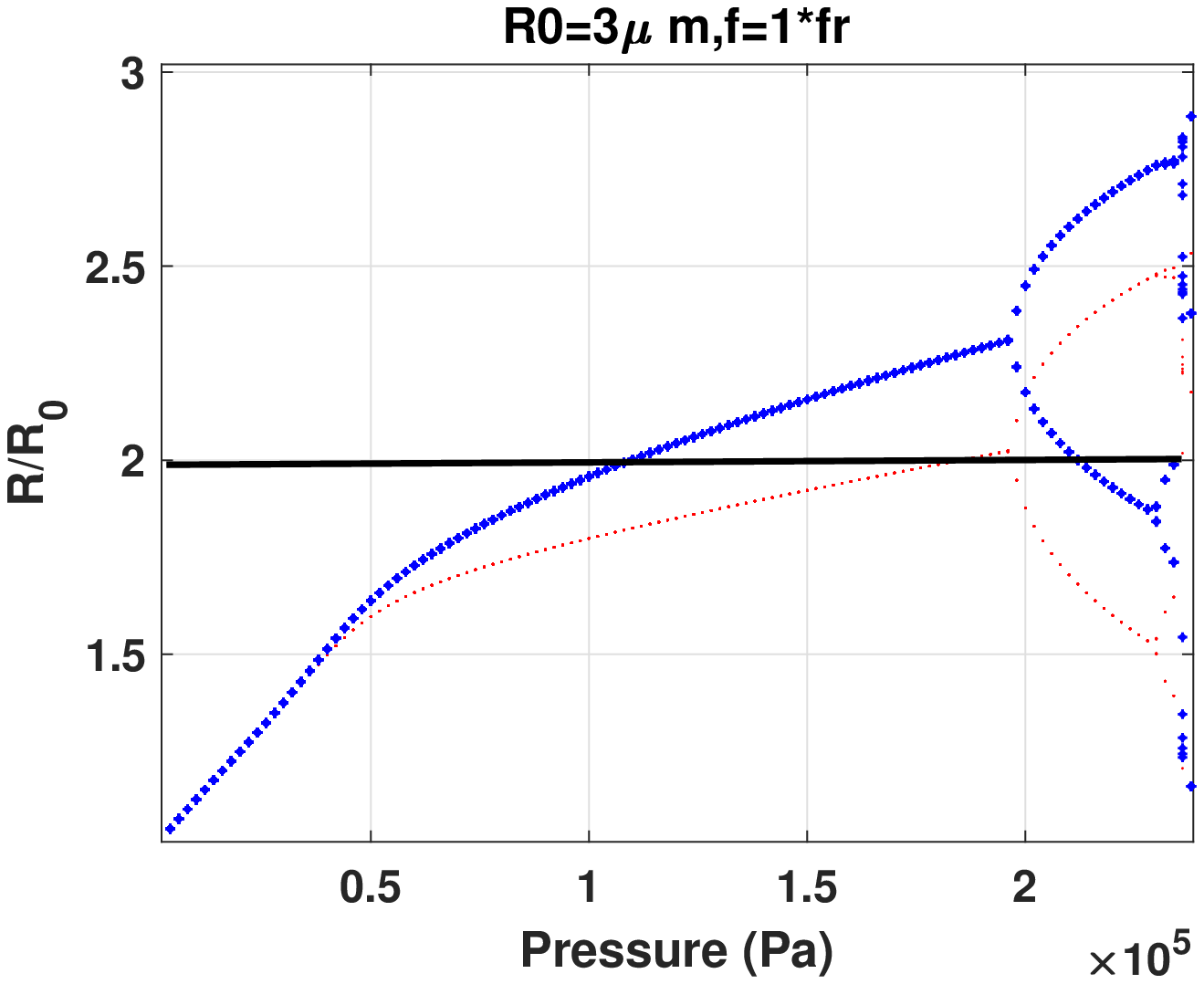}} \scalebox{0.33}{\includegraphics{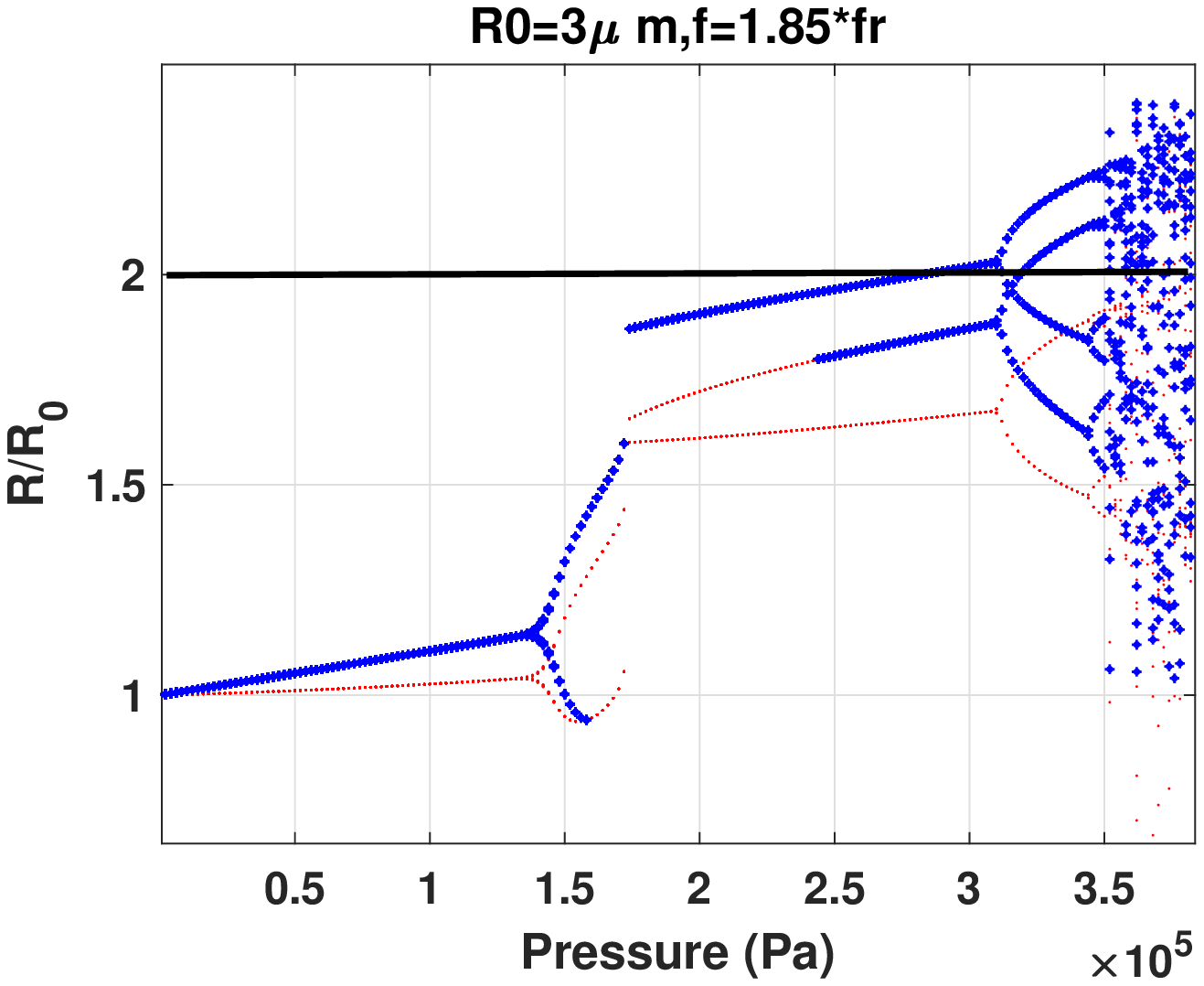}}\\
		\hspace{0.5cm} (c) \hspace{6cm} (d)\\
		\scalebox{0.33}{\includegraphics{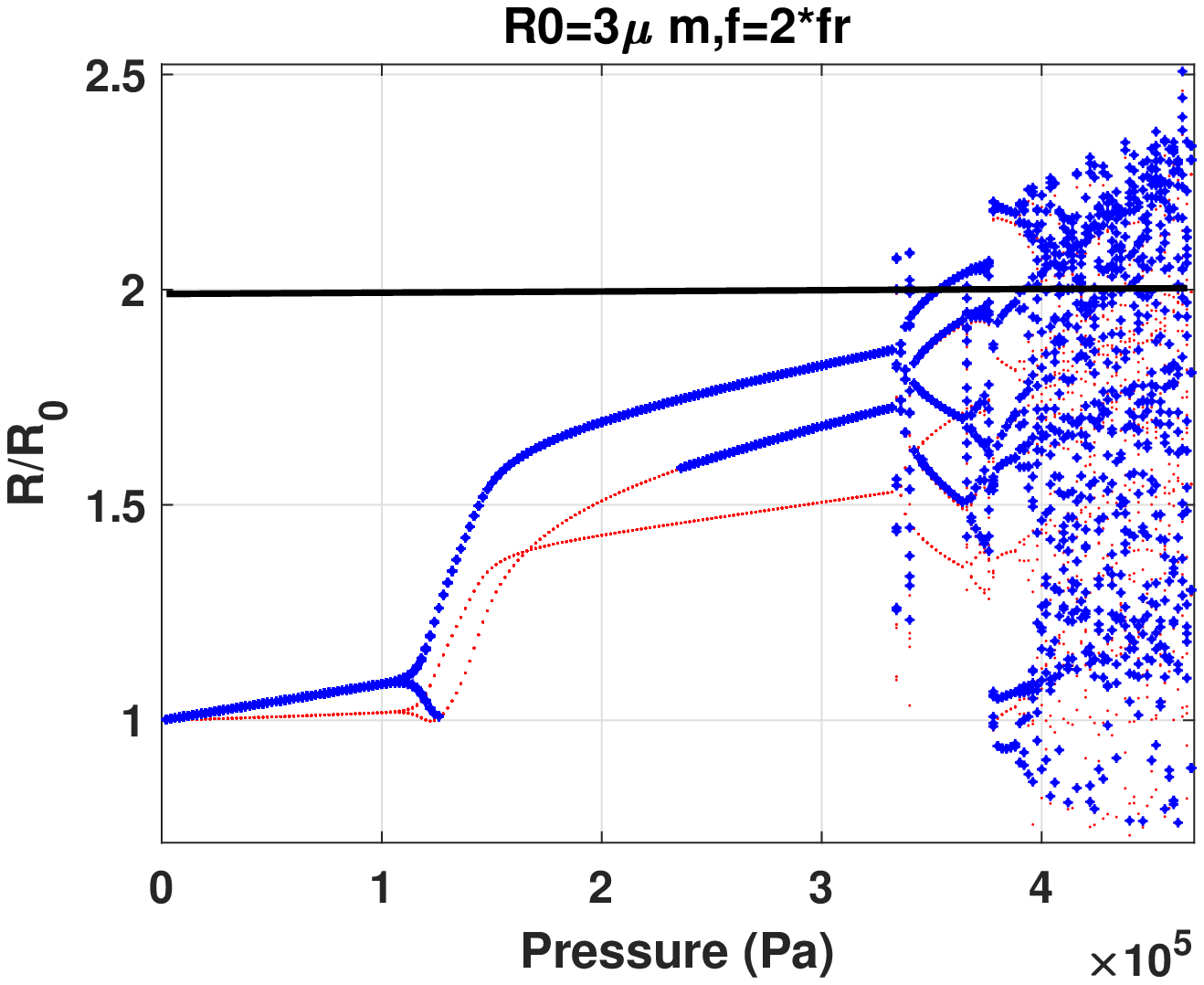}} \scalebox{0.33}{\includegraphics{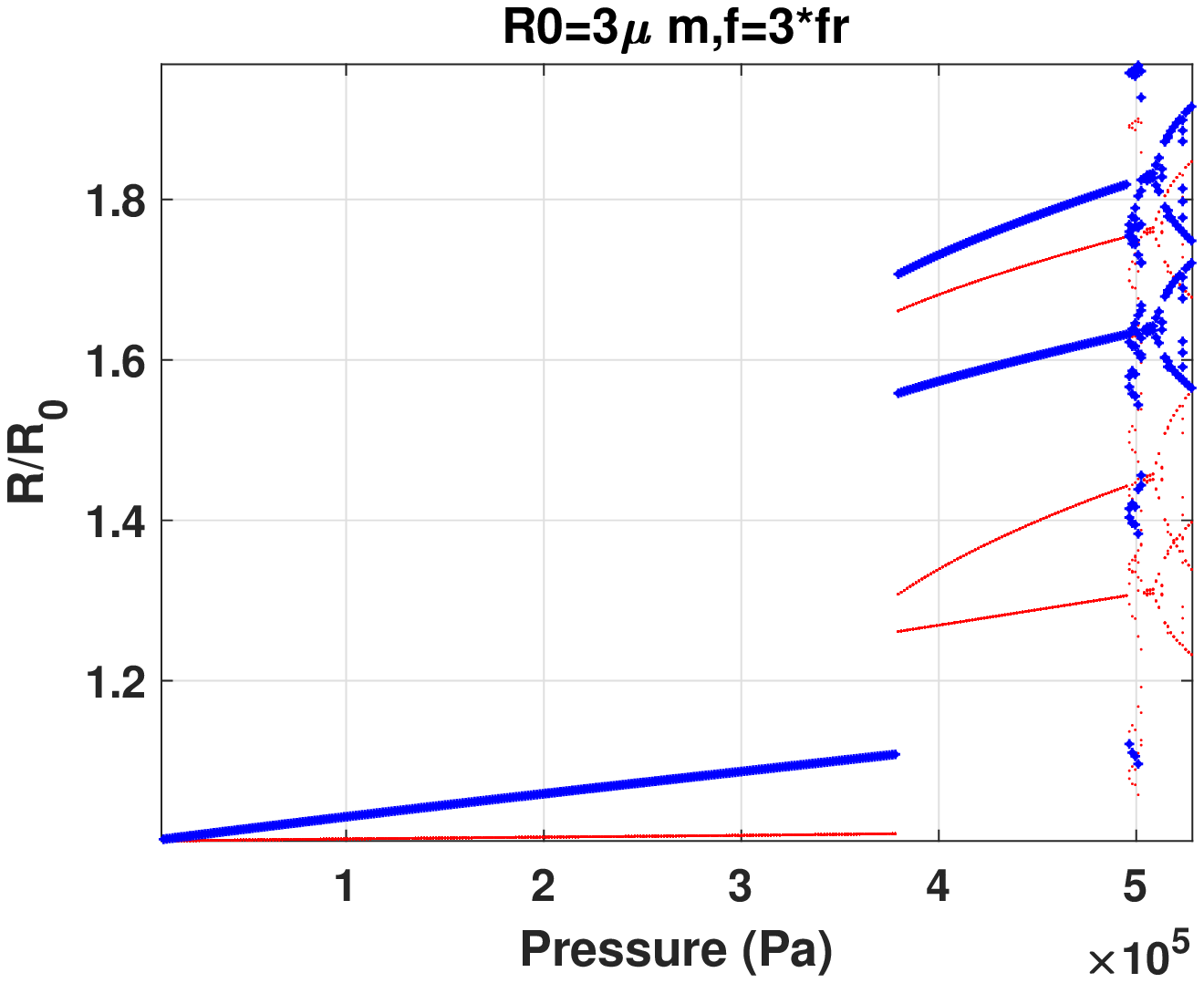}}\\
		\hspace{0.5cm} (e) \hspace{6cm} (f)\\
		\caption{Bifurcation structure of an air bubble with $R_0$= 3 $\mu$ m as a function of pressure (Red line is constructed using conventional method and blue line represents the method of peaks \cite{21}) a)f=0.35$f_r$ b) f=0.9$f_r$, c)f=$f_r$, d)f=1.85$f_r$, e)f=2$f_r$ and f)f=3$f_r$}
	\end{center}
\end{figure*}

\begin{figure*}
	\begin{center}
		\scalebox{0.33}{\includegraphics{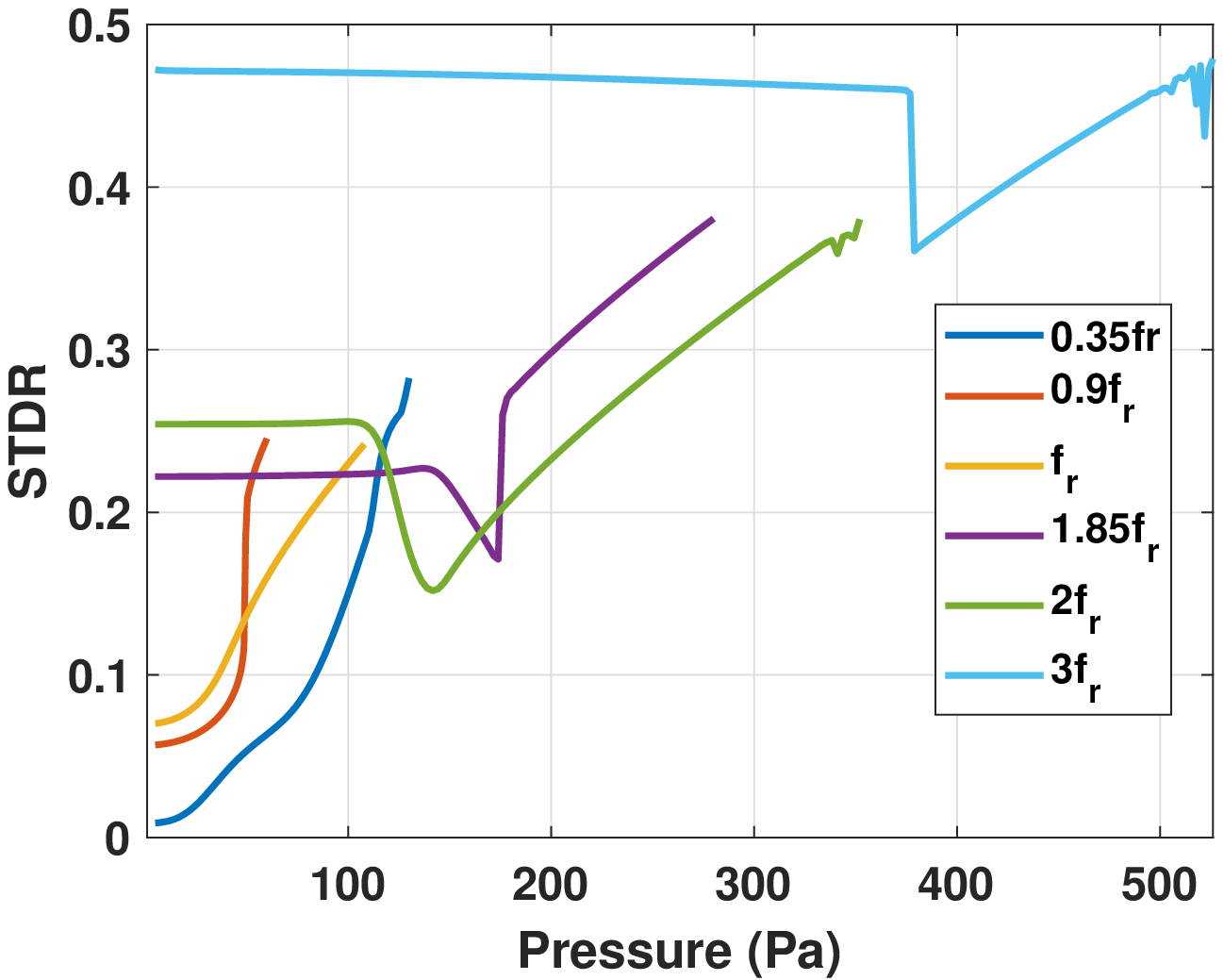}} \scalebox{0.33}{\includegraphics{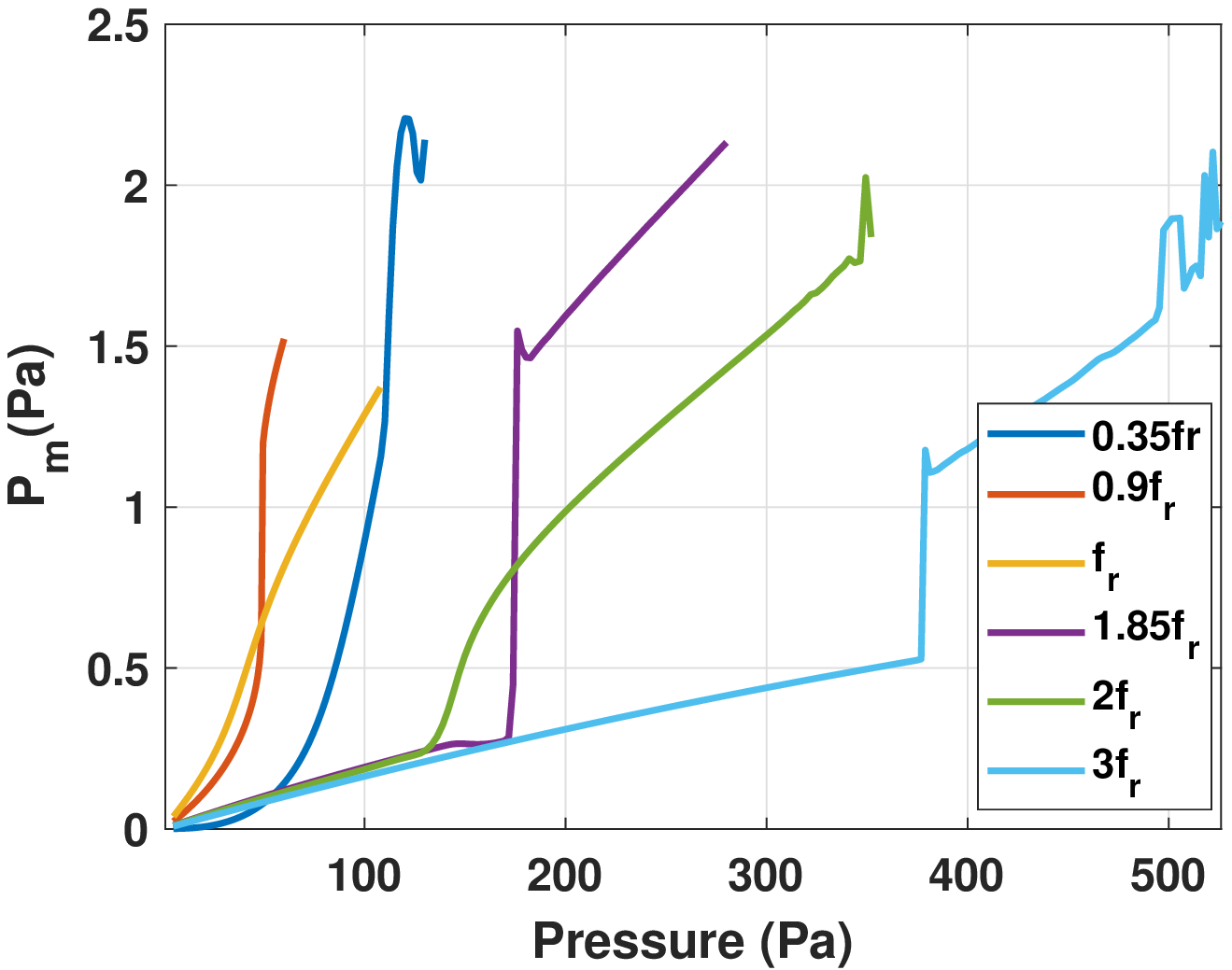}}\\
		\hspace{0.5cm} (a) \hspace{6cm} (b)\\
		\scalebox{0.33}{\includegraphics{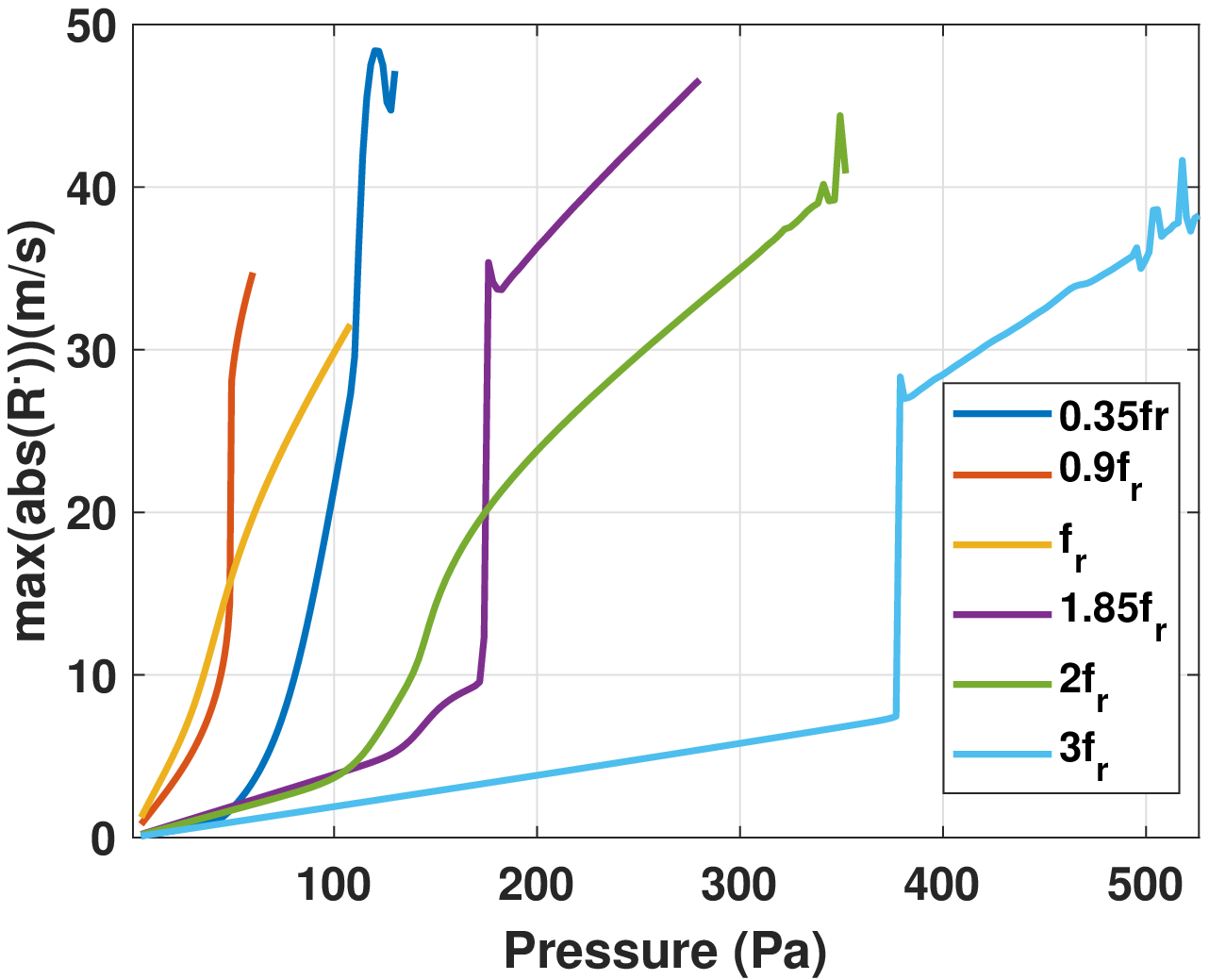}} 
		\scalebox{0.33}{\includegraphics{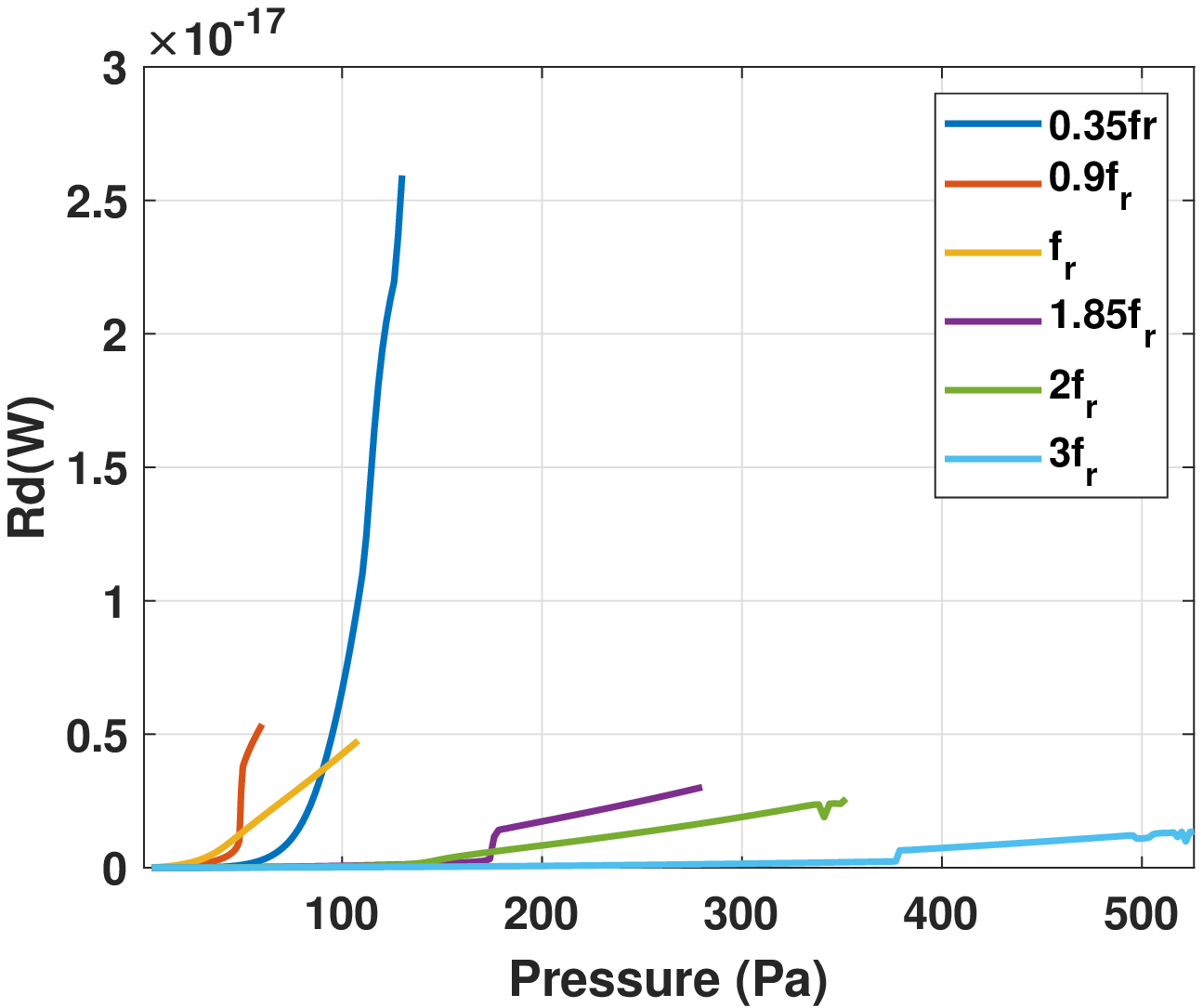}}\\
		\hspace{0.5cm} (c) \hspace{6cm} (d)\\
		\scalebox{0.33}{\includegraphics{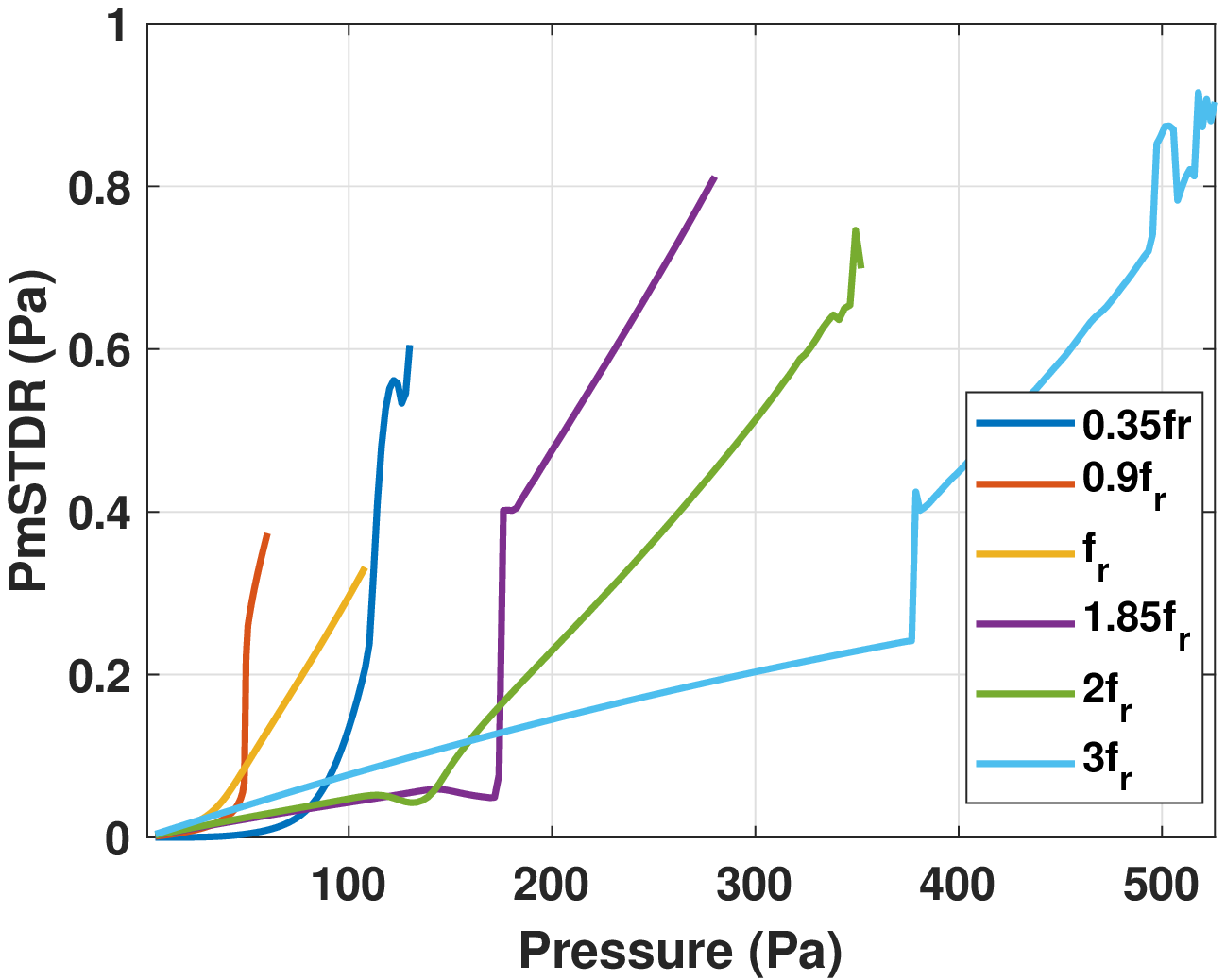}} 
		\scalebox{0.33}{\includegraphics{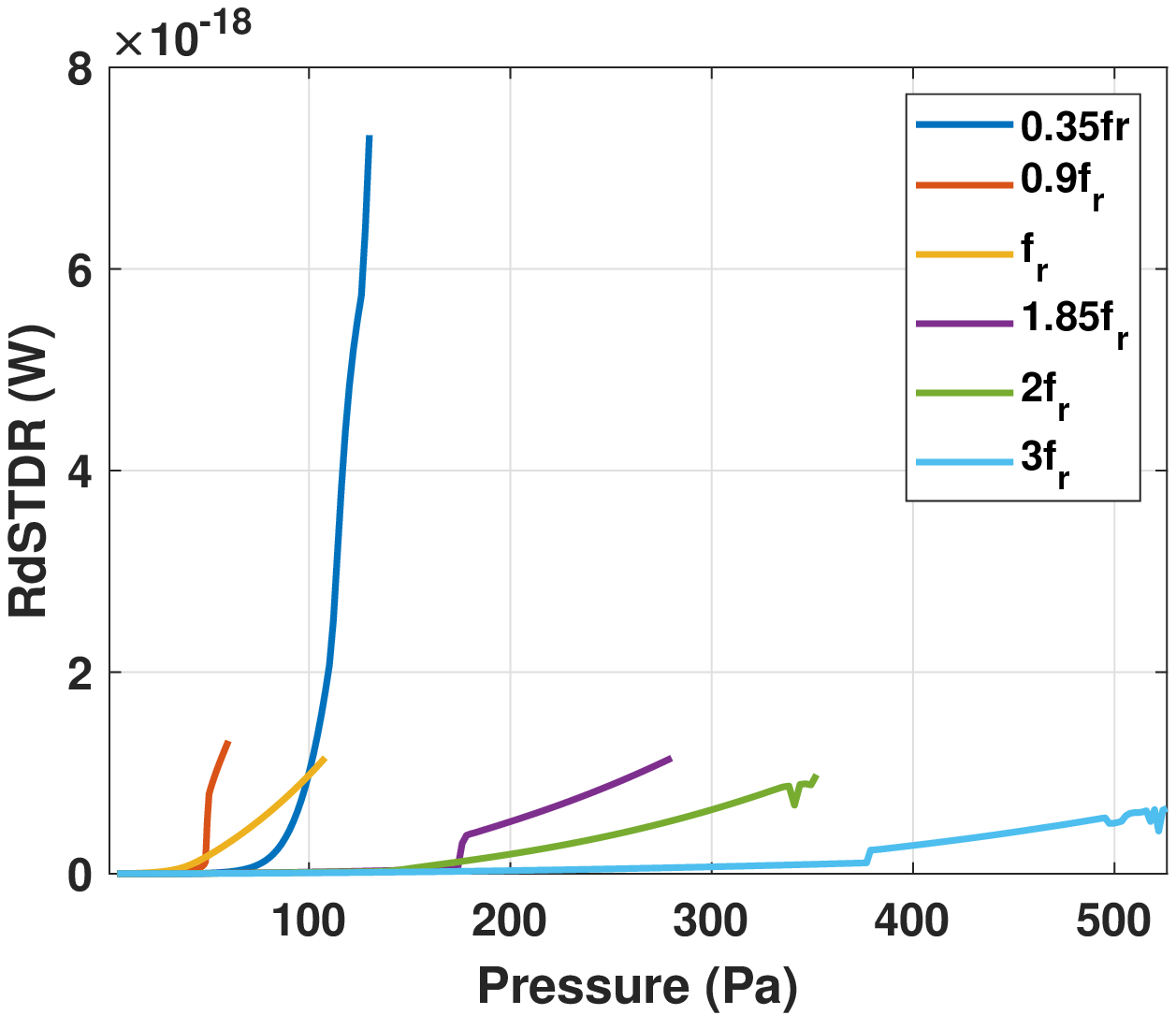}}\\
		\hspace{0.5cm} (e) \hspace{6cm} (f)\\
		\caption{a)STDR, b)maximum backscattered pressure amplitude$P_m$ and c) maximum absolute value of wall velocity d) Dissipation due to radiation Rd (acoustic power loss) e) PmSTDR and f)RdSTDR of an Air bubble with $R_0$=3 $\mu$m corresponding to the non-destructive oscillation regimes ( $\frac{R}{R_0}$ $\leq$ 2) in Fig. 2 .}
	\end{center}
\end{figure*}
The values for Rd, Td and Ld are pressure and frequency dependent. Rd grows with a faster rate than the other damping factors by pressure increase and there are regions in frequency and pressure domain in which Rd is stronger than other damping factors. This can have significant consequences for the optimization of applications. 
\subsection{Scattering to damping ratio (STDR)}
In this section, we attempt to define parameters to assess the efficacy of bubble related applications. Re-radiated (scattered pressure) by bubbles and attenuation \cite{1,2,3,4,5,6} are parameters that can be measured in real time and also can be calculated theoretically. Two characteristics typically define the efficacy of an application; enhanced bubble activity (e.g. scattering or microstreaming) and attenuation caused by the bubble activity.  For example, in ultrasound imaging, an Ultrasound contrast agent (UCA) with enhanced scattering properties is desired; however, if the same UCA causes significant attenuation in the medium, it significantly reduces imaging depth\cite{21}. Enhanced bubble activity is desired in applications; however, this enhancement should not be at the cost of increased attenuation that decreases the energy available to bubbles situated further in the beam path.\\ 
Scattered pressure by the bubbles is a function of bubble radial oscillations, wall velocity and acceleration; thus, there is a direct correlation between the intensity of bubble activity and the scattered pressure \cite{4}. Decreasing the attenuation of  bubbly media increases the power that can be delivered to bubbles and enhances the region of bubble activity. The scattering to attenuation ratio (STAR) was proposed in \cite{21} as a measure of the scattering effectiveness of contrast enhanced ultrasound. However, in their work and subsequent following studies, linear parameters were used to calculate STAR, thus the pressure dependence of STAR was not examined.\\STAR however is nonlinear and depend on the complex bubble dynamics. We observed in Fig. 1 that Rd can overcome the other damping factors for specific pressure and frequency ranges, and  thus there are potential parameter domains in which STAR can be maximized. Using the nonlinear formulation represented by Eq. 14 for the damping terms, we define a similar term to STAR which can also be used for nonlinear regimes of oscillations:
\begin{equation}
STDR= \frac{Rd}{Td+Ld+Rd}
\end{equation}
The nonlinear STDR is an important parameter in the characterization of bubbles in oceanography \cite{4,5,7,8} and ultrasound contrast agents shell spectroscopy \cite{6,9,10,11}. 
\subsection{Examination of main nonlinear oscillation regimes}
In order to have a better understanding of the effect of  nonlinear bubble oscillations on the STDR and to examine if this parameter can be used as a unifying parameter to assess application efficacy, we have considered 6 different oscillations regimes. Figure 2a-h shows the bifurcation structure of an air bubble with $R_0$=3 $\mu$ m as function of pressure as sonicated  by (2a) f=$0.35f_r$ ($f_r$ is the linear resonance frequency which is 1.0695 MHz here), (2b) f=$0.9 f_r$,  (2c) $f=f_r$, (2d) f=1.85$f_r$,(2e) f=2$f_r$,and (2f) f=3$f_r$ and f=3.65$f_r$. The red bifurcation is constructed using the conventional method of bifurcation analysis (method of peaks) and the blue curve is constructed using the method of maxima \cite{22}. We have shown that when the two bifurcations are plotted side by side we can reveal more intricate information about the dynamics of the system.\\Figure 2a shows that when bubble is sonicated with f=$0.35 f_r$ oscillations undergo 3rd order superharmonic resonance at $P_A\backsimeq$ 75 kPa (oscillations are of period one with 3 maxima \cite{22}) and when $P_A\backsimeq$ 115 kPa 3 period doublings (PD) occur resulting in 7/2 ultra harmonic (UH) resonance. Further pressure increase will likely result in bubble destruction at $P_A~130 kPa$ (since $\frac{R}{R_0}$ $>$ 2 \cite{22,23}). Fig. 2b shows that when bubble is sonicated with $f=0.9 f_r$ for $P_A ~< 200 kPa$ oscillations are of period 1 (P1) with one maxima; a saddle node bifurcation occurs at $P_A=50 kPa$ from a P1 with lower amplitude to a P1 with a higher amplitude. This is the result of sonication with pressure dependent resonance as is studied in detail in our previous work \cite{23}. Further pressure increase will result in bubble destruction before it can undergo period doubling. Figure 2c shows that when $f=f_r$ the oscillations are of P1 with one maxima and their amplitude grows as pressure increases. In this case as well, the pressure increase will likely result in bubble destruction before the occurrence of PD. When $f=1.85 f_r$ which is the pressure dependent SH resonance frequency \cite{24}  (Fig. 2d) P1 oscillations undergo a PD to P2 oscillations with two maxima at $\approxeq$ 145 kPa; and further pressure increase results in a saddle node bifurcation of the P2 oscillations with two maxima with lower amplitude to a P2 oscillations with one maxima of higher amplitude at~170 kPa. The bubble will likely to undergo destruction for $P_A>280 kPa$. This frequency is called the pressure dependent SH resonance frequency .  When the bubble is sonicated with $f=2f_r$ (linear SH resonance frequency \cite{25} (Fig. 2e)) P1 oscillations undergo a bow-tie shaped PD at~110 kPa, further pressure increase results in consecutive PDs leading to chaos and bubble destruction at $P_A  \backsimeq352$ kPa. When the bubble is sonicated with $f=3 f_r$ \cite{26} (Fig. 2f) P1 oscillations undergo period tripling through saddle node bifurcation at $P_A\backsimeq390 kPa$. Oscillations are of P3 with two maxima and bubble destruction occurs for a pressure region above $526\backsimeq kPa$.\\
\subsection{STDR and definition of a unifying parameter to assess the efficacy of applications}
In this section we will look at the dynamics of the scattering to damping ratio (STDR) in the nonlinear regimes of oscillations introduced in Fig. 2. To simplify the analysis we focus on oscillation regimes that result in stable and non-destructive bubble oscillation regimes ($\frac{R}{R_0}\leq 2$ \cite{22,23}). The STDR is calculated by integrating the damping parameters for the last 20 cycles of 60 cycle ultrasound pulses.\\
Figure 3a displays the STDR as a function of pressure. STDR is non-linear and pressure dependent. Sonication with $f=0.35f_r$ has the smallest STDR, however by increasing the incident pressure it becomes stronger than STDR in case of sonication with $f=f_r$ and $0.9f_r$.
For sonications above resonance, the STDR is higher for higher frequencies. Sonications with $f=3f_r$ results in the maximum STDR of $\sim$ 0.49. When SH oscillations occur, STDR undergoes a fast decrease but quickly recovers and can reach higher values. The higher value for STDR at higher frequencies is due to the increase in scattering and a decrease in thermal damping which is the most dominant damping factor for lower frequencies. The concomitant decrease of STDR  with SH oscillations is due to the increase in  the thermal damping due to higher amplitude oscillations.\\
Although higher STDR indicates the higher ratio of scattering to the total damping by bubble oscillations; a higher STDR does not necessarily equal to enhanced bubble activity. For example, despite the highest value for STDR when $f=3f_r$, the bubble has minimum oscillation amplitude before the occurrence of P3, meanwhile the STDR is 2-5 times smaller for the case of sonication with resonance frequency despite higher oscillation amplitudes. This suggests that knowledge of STDR by itself is not sufficient for use in application optimization.
STDR can have a large value if the ratio of scattering to total damping is high, but at the same time if the scattering is negligible, bubble oscillations may not result in any signal enhancement in ultrasound images or any tangible drug delivery.\\  
Figs. 3b-d respectively show the maximum backscattered pressure ($P_m$), maximum absolute bubble wall velocity ($|\dot{R}|_{max}$) and damping due to pressure re-radiated by bubbles (Rd).  $P_m$ is a good measure for the maximum instantaneous bubble activity (a higher value of $P_m$ correlates to higher echogenecity) and higher Rd means higher dissipated energy (for example a higher Rd is more desirable in bubble enhanced heating \cite{27,28} where increased absorption due to re-radiation of superharmonics enhances tissue heating). For pressures below 100 kPa, Figs. 3a-d show that $P_m$ is maximum when $f=0.9 f_r$ and $f=f_r$. However, for this pressure range the highest STDR occurs at the highest frequencies ($STDR_{3f_r}>STDR_{2f_r}>STDR_{1.85f_r}$). $STDR_{3f_r}$ is 2-5 times larger than $STDR_{f_r}$ (for an instance $STDR_{3f_r}(100kPa)\sim0.49$ and $STDR_{f_r}(100kPa)\sim0.22$). For $f=0.35f_r$, the occurrence of the 3rd harmonic resonance is concomitant with a steep increase in $STDR_{0.35f_r}$; at the same time $P_m$, $|\dot{R}|_{max}$ and Rd become stronger than all the other frequencies studied here (as an example ${P_m}$ of $0.35f_r$ reaches a maximum of 2.5 Pa at $\sim$127 kPa while $P_m$ of ${f_r}$ reaches a maximum of 0.5 Pa). Maximum energy dissipation due to Rd occurs for sonication below resonance at $f=0.35f_r$ and $f=0.9f_r$ with the maximum dissipation at $f=0.35f_r$ being 4.8 times higher than the maximum dissipation at $f=0.9f_r$.\\
In order to incorporate the maximum bubble activity and STDR we define a new parameter which is the result of the multiplication of $P_m$ by STDR: 
\begin{equation}
PmSTDR= \frac{P_m*Rd}{Td+Ld+Rd}
\end{equation}
This parameter can be used to find parameter regions that result in high bubble  activity while reducing the effect of damping. One application would be the optimization of contrast enhanced ultrasound by increasing the echogencity and decreasing the attenuation. Figure 3e shows the PmSTDR as a function of pressure for non-destructive regimes of oscillations. For the parameter ranges used, sonication with f=$1.85f_r$ (pressure dependent SH resonance) and f=$3f_r$ (3rd order SH resonance) resulted in the maximum PmSTDR. The maximum possible PmSTDR at f=$3f_r$ is 3 times higher than the maximum PmSTDR at $f=f_r$.\\
In some applications like bubble enhanced heating in HIFU the goal is to maximize the absorbed power that is re-radiated by bubbles while minimizing the absorption of re-radiated energy by bubbles outside of the focal region \cite{27,28}. Thus in order to capture both the damping effect and the absorption due to re-radiation we can multiply STDR by Rd: 
\begin{equation}
RdSTDR= \frac{Rd*Rd}{Td+Ld+Rd}
\end{equation}
Figure 3f plots RdSTDR as a function of the acoustic pressure and for non-destructive regime of oscillations. RdSTDR has a maximum for $f=0.35f_r$ which is 6.65 times larger than the maximum RdSTDR at resonance. The higher frequency component in the scattered signal at 3rd harmonic resonance when $f=0.35f_r$ leads to higher absorption of the scattered pressure in tissue as higher frequencies have larger attenuation. Increase in the driving frequency decreases the RdSTDR with maximum RdSTDR at f=$3f_r$ is 1.8 times smaller than RdSTDR at resonance. 
\section{Discussion}
Existence of bubbles increases the attenuation of the host media. Understanding the phenomena related to bubble dynamics and the optimization of the bubble activity requires a solid understanding of the changes of the medium attenuation due to bubble pulsations. The attenuation of the medium is a function of the bubble dynamics which are shown to exhibit nonlinear beahvior including subharmonic, superharmonic and chaotic oscillations \cite{22,23,24,25,26,29,30,31}.
The majority of the previous studies considered linear approximations (small bubble amplitude (e.g. \cite{4,5,6,7,8})) to infer the changes of the attenuation of the medium or sound speed \cite{5,6,7,8,32} or semi-linear approaches where only the nonlinear changes of the scattering cross section were accounted for in the equations (e.g. \cite{9}). However, bubble oscillations are nonlinear in the majority of applications (e.g. \cite{27,28,32,33,34,35,36}). In these applications, the higher acoustic pressures result in non-linear large amplitude bubble oscillations; thus, linear or semi-linear approximations fail to accurately model the medium attenuation. Moreover, bubble oscillations can significantly be influenced by thermal effects \cite{37,38,39,40}; however thermal effects are largely neglected or simplified using linear approximations \cite{5,6,7,8,9,37,40,41}.\\
To accurately understand the attenuation of the medium, effects of large amplitude oscillations of the bubbles and their dependence to the local acoustic pressure should be incorporated in the model. Furthermore, nonlinear effects of liquid comprehensibility and thermal damping should be included.\\
Louisnard \cite{1} used the conservation of momentum and energy in a bubbly medium and developed nonlinear damping terms for thermal damping (Td) and liquid viscous damping (Ld). He has shown that the attenuation is pressure dependent and as pressure increases predictions of the linear model become invalid (e.g. orders of magnitude less than nonlinear model). Jamshidi and Brenner \cite{2} incorporated the effects of the compressibility of the medium to the first order of Mach number and derived the nonlinear terms for the radiation damping, thermal and viscous damping. We corrected this model and showed that the predictions of the corrected formulations are in excellent agreement with the predictions of the acoustic pressure power theory.\\ 
It was shown that Rd, Td and Ld are nonlinear and depend on pressure and frequency. Rd is the weakest dissipation mechanism at lower pressures (e.g. $P_A<20 kPa$). As pressure increases, Rd grows with a faster rate compared to Td and Ld and at specific frequencies (e.g. pressure dependent resonance frequency \cite{23}) becomes the dominant dissipation effect. Thus, comprehensibility effects become important even at moderate pressures (e.g. 100 kPa (Fig. 1h)). This shows that models for ultrasound contrast agents (UCAs) \cite{6} that neglect or simplify liquid compressibility effects may loose accuracy even at moderate pressures.\\
Increased attenuation of ultrasonic waves due to the presence of bubbles limits the delivery of sufficient energy to activate bubbles that are at the focus and/or limits the regions of enhanced bubble activity. In therapeutic applications of ultrasound (e.g. drug delivery \cite{36}, high intensity focused ultrasound \cite{26}) the increased attenuation of pre-focal bubbles may result in undesirable heating in the healthy tissue. In diagnostic applications of ultrasound, increased attenuation of bubbles in the beam path creates shadowing effects that deteriorate the images of underlying tissues \cite{21}. In sonochemical reactors, inhomogeneous pressure distribution inside reactors from pressure dependent attenuation reduces the yields efficacy. Accurate formulations for nonlinear damping due to bubbles significantly assists in understanding the mechanism of nonlinear attenuation and aids in designing protocols that minimizes the unwanted effects. Furthermore, knowledge on the scattered (re-radiated) energy from the bubbles can be used as a measure of the bubble activity. Thus, to this end the scattering to damping ratio (STDR), PmSTDR and RdSTDR can be used as a complete set of parameters to assess the efficacy for specific applications. Accurate calculation of PmSTDR or RdSTDR can help in designing the optimized frequency and pressure ranges to reduce the unwanted pre-focal attenuation and enhance the bubble activity in focus. For example Fig. 3f shows that $RdSTDR_{0.35f_r}$ is below $RdSTDR_{f_r}$ for pressures below 100 kPa (e.g. 3.8 times smaller at 90 kPa); as pressure increases above 100 kPa $RdSTDR_{0.35f_r}$ becomes 6.65 times larger than the case of sonication with $f_r$. By taking advantage of the steep pressure gradients of focused transducers and setting the focal pressure slightly above 100 kPa, this property can be used to reduce the absorption of ultrasound in pre-focal region while maximizing the absorption and bubble activity in the target focal region \cite{42,43}.
\section{Conclusion}
This study provides accurate formulations for the mechanisms of power dissipation during propagation of the ultrasonic waves through a bubbly medium. The approach used in this paper can be applied to derive equations of power loss in other types of media like sediment \cite{3,4,7,8} or tissue \cite{44,45} and encapsulated bubbles. Application of the nonlinear formulations provide accurate pressure dependent predictions in studies related to characterization of bubbles in underwater acoustics \cite{3,4,7,8} and characterization of shell parameters of encapsulated bubbles \cite{9,10,46,47,48,49,50}. Moreover, the exposure parameters of the applications can be optimized to enhance a particular effect of interest.
\section{Acknowledgments}
The work is supported by the Natural Sciences and Engineering Research Council of Canada (Discovery Grant RGPIN-2017-06496), NSERC and the Canadian Institutes of Health Research ( Collaborative Health Research Projects ) and the Terry Fox New Frontiers Program Project Grant in Ultrasound and MRI for Cancer Therapy (project $\#$1034). A. J. Sojahrood is supported by a CIHR Vanier Scholarship.

\end{document}